\documentclass[proceedings, preprint]{rmaa}

\usepackage{paralist}

\usepackage{psfrag,color}
 
\SetYear{2022}
\SetConfTitle{Prospects for low-frequency radio astronomy in S. A.}

\title{The Instituto Argentino de Radioastronomía (IAR): Past, present, and future } 

\author{
  G. E. Romero\altaffilmark{1,2} 
  }

\altaffiltext{1}{Instituto Argentino de Radioastronom\'{\i}a (IAR), Argentina (romero@iar.unlp.edu.ar).}

\altaffiltext{2}{ Facultad de Ciencias Astron\'omicas y Geof\'{\i}sicas, Universidad Nacional de La Plata, Argentina (romero@fcaglp.unlp.edu.ar).}

\shortauthor{Romero}
\shorttitle{IAR: Past, present, future}

\listofauthors{G. E. Romero}
\indexauthor{Romero, G. E.}
\abstract{I present a brief review of the history of the Instituto Argentino de Radioastronomía, a description of its current facilities and projects, and a view of his prospects for the future.}

\resumen{Presento un breve relato de la historia del Instituto Argentino de Radioastronomía, junto con una descripción de sus instalaciones y proyectos actuales, así como una visión de sus perspectivas futuras.}
\addkeyword{History of astronomy}
\addkeyword{Instrumentation: detectors}
\addkeyword{Instrumentation: interferometers}
\addkeyword{Radio continuum: general}
\addkeyword{Radio lines: general}

\begin{document}
% Typeset article header
\maketitle

\section{Origins}
\label{sec:origins}

\begin{figure}[!t]
  \includegraphics[width=\columnwidth]{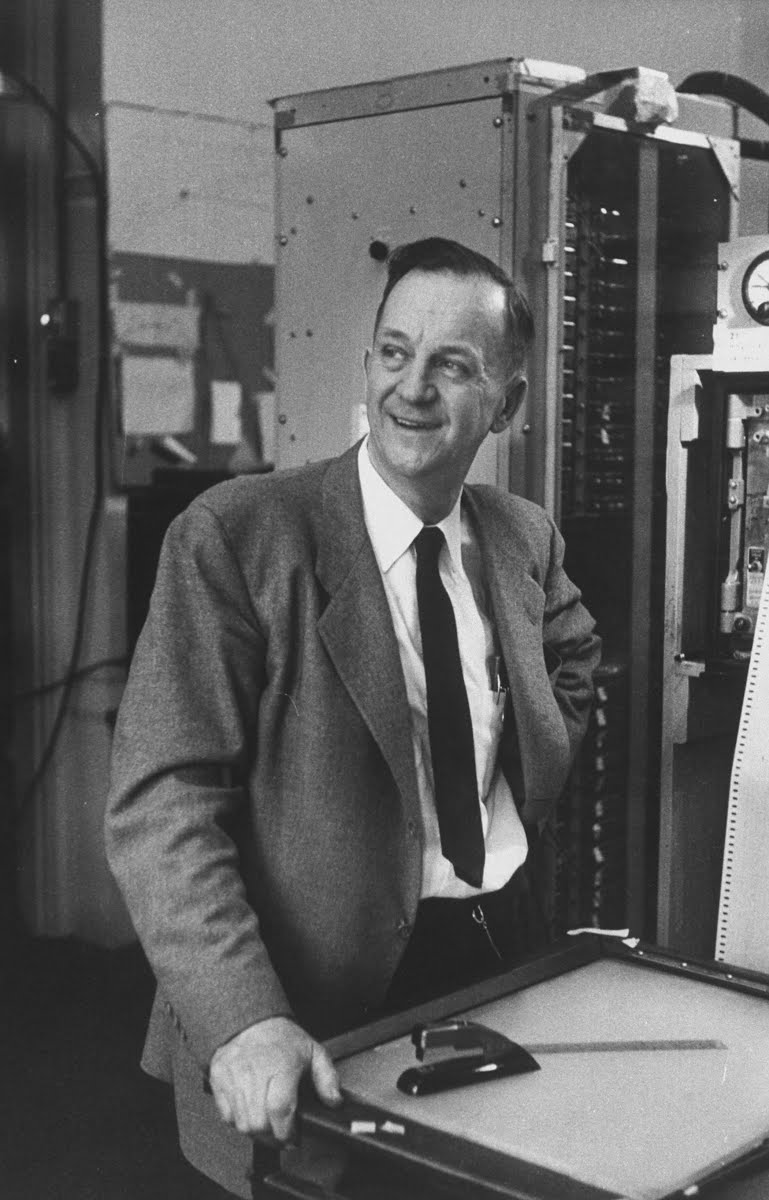}
  \caption{Merle Anthony Tuve. Director of the Department of Terrestrial Magnetism (DTM) at the Carnegie Institution for Science (1946–66) and key person in the development of radio astronomy in Argentina.}
  \label{fig:Tuve}
\end{figure}

The origins of radio astronomy in South America and of the Instituto Argentino de Radioastronomía (IAR), the first observatory devoted to this branch of astronomy in this part of the world, are associated with the name of Merle Anthony Tuve (June 27, 1901 - May 20, 1982, Figure \ref{fig:Tuve}). He was an American geophysicist who pioneered the use of pulsed radio waves. His discoveries paved the way for the development of radar and nuclear physics. Merle Tuve was director of the Department of Terrestrial Magnetism (DTM) at the Carnegie Institution for Science (1946-66), and from that position he played a key role in the creation of the IAR.

Tuve became interested in the nascent field of radio astronomy when he learned in 1952 of the first detection of the HI line from the galactic plane by Edwin and Purcell of Harvard University. He soon built a radiometer and began his research in radio astronomy. He soon realized the importance of having radio observatories in the southern hemisphere, since most of the Galaxy is visible only from that part of the world. At the time, the southern sky was virtually unexplored in the radio spectrum. Since Australia was already working on the construction of what would become the Parkes Observatory, Tuve thought that South America would provide an excellent platform for the installation of a second southern radio observatory. It would be more easily accessible from the United States and would be an important complement to observational studies to be conducted from Australia. In 1957 he visited several countries in the region: Brazil, Argentina, Chile, and Peru. In Argentina, astronomy had been systematically developed since President Domingo Faustino Sarmiento founded the Córdoba Observatory in 1871, and Tuve's proposal was enthusiastically received. As a direct consequence of Tuve's visit and his promise to send the country the necessary components to build a solar interferometer, the University of Buenos Aires (UBA) created, on November 13, 1958, the Commission of Astrophysics and Radio Astronomy (CAR), integrated by Dr. Félix Cernuschi, Dr. Enrique Gaviola and Eng. Humberto Ciancaglini, with Dr. Gaviola as its president.

Gaviola (1900-1989) was one of Argentina's most distinguished scientists.  He received his doctorate under the supervision of Max von Laue in Berlin. Recommended by Einstein, he received a fellowship from the Rockefeller Foundation to continue his research in the United States. Between 1928 and 1929 he worked at the DTM, where he met Tuve. 

Tuve sent the parts of the interferometer to the CAR, so that the construction of the different parts was done in the offices of the Commission, and then the instrument was installed on the land of the Faculty of Agronomy of the UBA. It consisted of 16 antennas operating at 86 MHz. It did not produce any scientific results, but it was important for the training of students and engineers involved in the project. At the same time, other efforts were made to train some engineers in the new technologies: Jorge Sahade, who was in the USA at the time, and the then director of the National Radio Astronomy Observatory (NRAO), Dr. Otto Struve, organized a trip to the DTM and NRAO for two young engineers: Emilio Filloy and Rubén Dugatkin. They were supported by grants from the Scientific Research Commission (CIC) of the Province of Buenos Aires, Argentina, and finally traveled to the United States in September 1961.

On December 7, 1961, Tuve sent a letter to Dr. Bernardo Houssay, President of the National Research Council of Argentina (CONICET), proposing the establishment of a radio astronomy station in Argentina, along with a memorandum describing the main instrument of the facility: a 30-m single-dish equatorial mount radio telescope. The instrument would be able to cover from the south celestial pole to $-10\deg$ in declination and from $-2$ to $+2$ hours in right ascension. The DTM would provide all materials for the construction of the telescope. This letter from Tuve initiated a series of exchanges with Houssay that would eventually lead to the creation of the IAR.

\section{History}
\label{sec:history}

On April 27, 1962, CONICET created the National Institute of Radio Astronomy (INRA). Soon after, an agreement was signed with the UBA, the University of La Plata (UNLP), and the CIC, stating that all these institutions would work together to support the new institute. Finally, the name was changed to Instituto Argentino de Radioastronomía (Argentine Institute of Radio Astronomy) to avoid confusion with another institute dedicated to agricultural technologies, but with the same acronym. 

The history of the IAR has been presented in detail by Bajaja (2009). What is offered in this section is a brief account along with some personal views. 

\begin{figure}[!t]
  \includegraphics[width=\columnwidth]{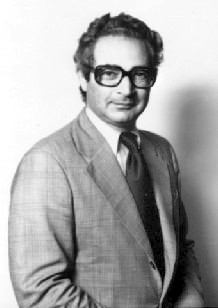}
  \caption{Carlos Varsavsky, first director of IAR.}
  \label{fig:Varsavsky}
\end{figure}

Dr. Carlos M. Varsavsky (1933-1983, Figure \ref{fig:Varsavsky}), from UBA, was chosen as director of the Institute and Dr. Carlos Jaschek of the La Plata Observatory was appointed as deputy director. Eng. Juan del Giorgio was appointed technical advisor. Varsavksy received his Ph.D. in astronomy from Harvard in 1959. He was known as the translator into English of the classic book \textit{Cosmic Radio Waves} by I.S. Shklovsky. In 1960 he returned to Argentina and became Professor of Physics at the University of Buenos Aires, where he lectured on radio astronomy and joined the CAR group. The original staff of the IAR consisted of the aforementioned authorities, the engineers Rubén Dugatkin, Emilio Filloy (both recently returned from the USA), Omar González Ferro, and the students Valentín Boriakoff (electronics), Fernando Raúl Colomb, and Esteban Bajaja (both students of physics).

\begin{figure}[!t]
  \includegraphics[width=\columnwidth]{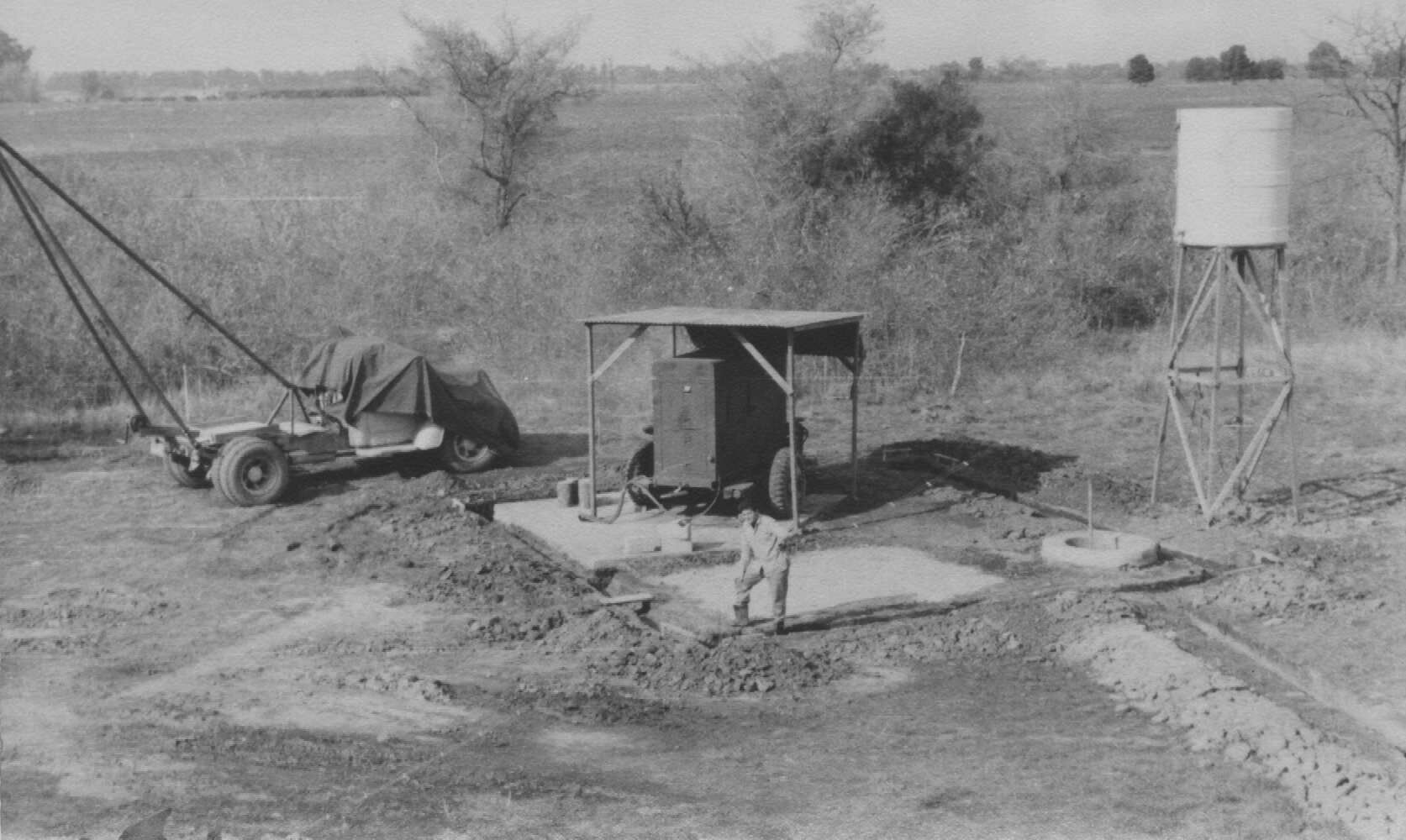}
  \caption{Construction of the first workshop at IAR's estate. The International truck and the original Diesel generator can be seen, along with the first water tank. }
  \label{fig:first-facilities}
\end{figure}

The first task of the group was to find a suitable site for the installation of the radio telescope. The site should be accessible from both La Plata and Buenos Aires, but relatively isolated to minimize radio interference from human sources. The selected location was within the Pereyra Iraola Park, a protected reserve located near Villa Elisa, 40 km from the city of Buenos Aires. Ten hectares were ceded by the Province of Buenos Aires through an agreement signed on October 30, 1962. Almost immediately, work began on the site. A 1944 International truck and a 50 kVA ex-military diesel generator were used to build the first workshop and other facilities (see Figure \ref{fig:first-facilities}). Amazingly, both are still in use in 2023.

\begin{figure}[!t]
  \includegraphics[width=\columnwidth]{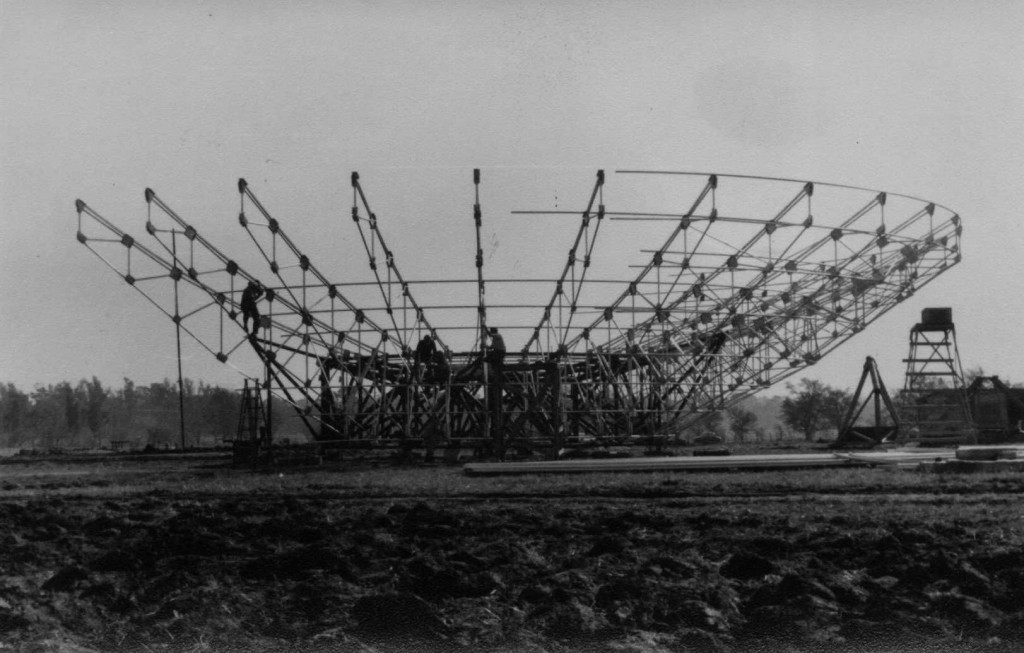}
  \caption{Construction of the dish of the first radio telescope at the IAR. Early 1964. }
  \label{fig:a1-01}
\end{figure}

To supervise the work of assembling the radio telescope, at the beginning of November 1963, Eng. Everett Ecklund was sent from the Carnegie Institution of Washington (CIW). The construction started on November 14th. The construction process is described in detail by Bajaja (2009) and will not be repeated here. Figure \ref{fig:a1-01} shows the construction of the dish, still resting on the ground.

\begin{figure}[!t]
  \includegraphics[width=\columnwidth]{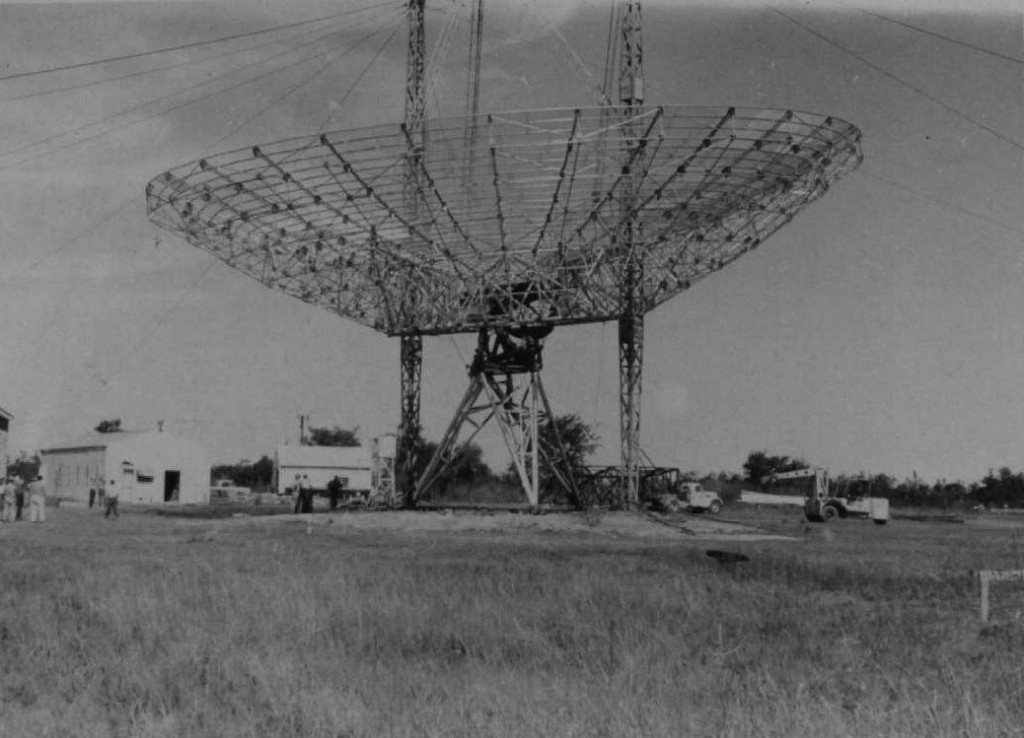}
  \caption{Mounting the radio telescope disk on its pedestal. 1965. }
  \label{fig:a1-02}
\end{figure}

Once the dish was completed, a first receiver with 10 channels was installed and on April 1, 1965, the first detection of HI was achieved. The dish (central ring + platform + ribs + surface) was supported on pillars on the ground, facing the zenith. This historic landmark occurred only 12 years after the original detection of the HI line at Harvard. It was a milestone in the history of the IAR and radio astronomy in the southern hemisphere. At the end of that same year, the dish was mounted on the pedestal (Figure \ref{fig:a1-02}).

\begin{figure}[!t]
  \includegraphics[width=\columnwidth]{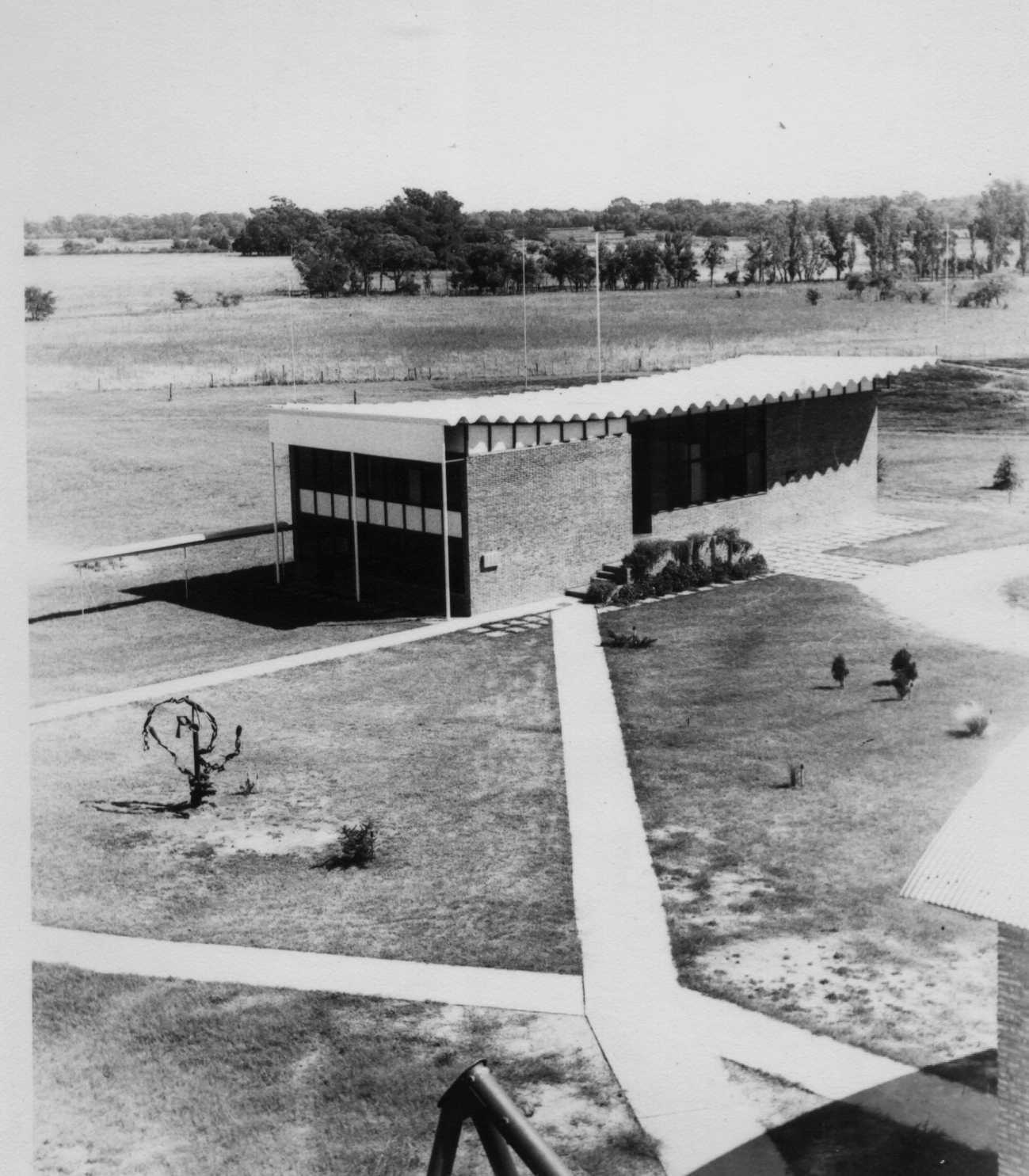}
  \caption{Main building of IAR, finished in early 1966.}
  \label{fig:principal}
\end{figure}

\begin{figure}[!t]
  \includegraphics[width=\columnwidth]{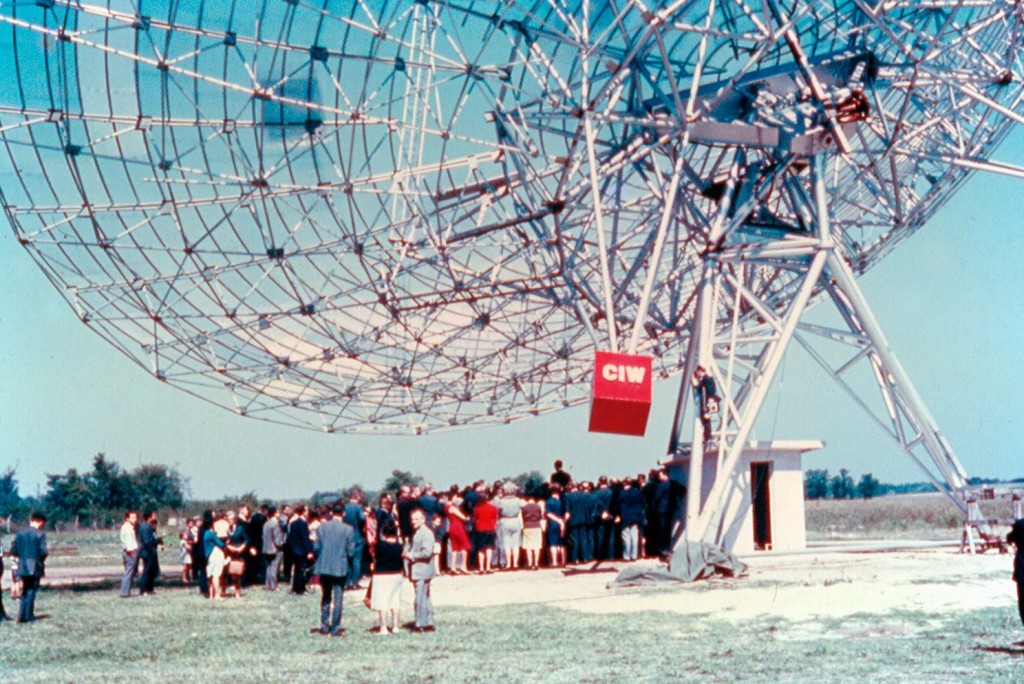}
  \caption{March 26, 1966: inaugurating ceremony of the first radio telescope of IAR.}
  \label{fig:inaguracion}
\end{figure}

As the construction of the radio telescope progressed, other buildings were added to the complex: a main building for offices, library, and control of the instrument (see Figure \ref{fig:principal}); and a building with laboratories and workshops for electronics.  Finally, the inauguration of the radio telescope took place on March 26, 1966, in the presence of personalities from 
national and foreign institutions, including H. Van de Hulst, the first proponent of the HI 21-cm line as a diagnostic tool of the interstellar medium (Figure \ref{fig:inaguracion}). During the inauguration ceremony, speeches were given by Dr. Tuve, in his capacity as Director of the DTM and representative of the CIW, and by Dr. Varsavsky, as Director of the IAR. Soon after, the scientific observations began: IAR as a radio astronomy observatory was born.

It did not take long for politics to interfere with the normal development of activities. On June 28, 1966, the Argentine Armed Forces carried out a coup against the constitutional government of Arturo Umberto Illia and established the self-proclaimed ``Argentine Revolution''.  On July 29, 1966, five academic departments of the University of Buenos Aires (UBA), occupied by students, professors, and graduates, were brutally repressed and evacuated in opposition to the decision of the \textit{de facto} government to intervene in the universities and overthrow the government regime. The episode in which professors were beaten and expelled was known as ``The Night of the Long Batons.'' Subsequently, the military government of General Juan Carlos Onganía decided to unilaterally revoke the academic freedom established in the 1918 university reform. Most professors resigned and many left the country.

\begin{figure}[!t]
  \includegraphics[width=\columnwidth]{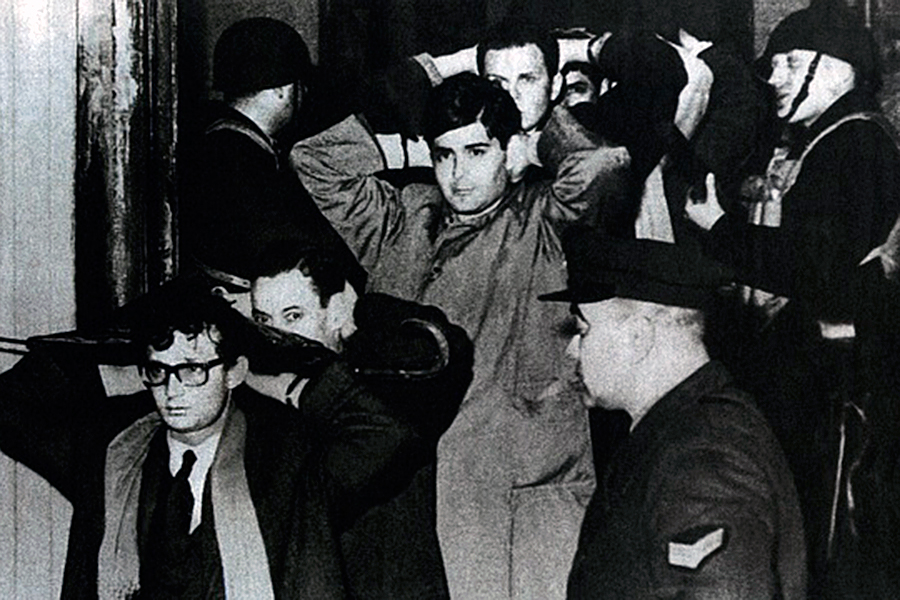}
  \caption{July 29, 1966: ``Night of the Long Batons'', at the Deparment of Physics, UBA. Varsavsky, wounded in the head, can be seen first in the row, with an umbrella.}
  \label{fig:bastones}
\end{figure}

\begin{figure*}
  \centering
  \includegraphics[width=0.80\textwidth]{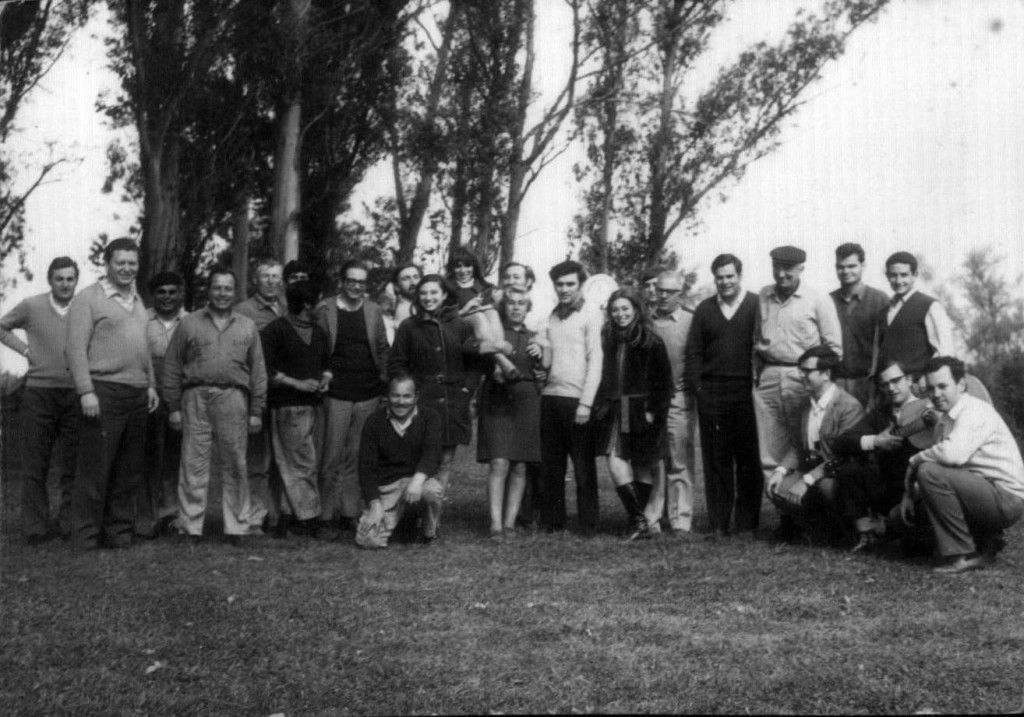}
  \caption{IAR staff toward 1966. }
  \label{fig:staff1966}
\end{figure*}

Varsavsky, who was wounded in the episode (Fig. \ref{fig:bastones}), resigned from the university and moved his office to the IAR. Many students came with him. Some of them later became famous scientists: Diego and Catherine Cesarsky, Jorge Vernazza, Peter M\'esz\'aros, Federico Strauss, Zulema Abraham, Silvia Garzoli, Esteban Bajaja, Raul Colomb, and Wolfgang Poeppel, among others (see Fig.~\ref{fig:staff1966}).

Despite the political turmoil, research with the new radio telescope continued. Then, in 1969, during the month of May, some serious incidents occurred in the cities of Rosario and Corrientes, with the death of two students. In the large city of Córdoba, some unions and groups of students called for a strike to be held on May 30. Then, on May 28, a note signed by the staff of the IAR, criticizing the regime of the de facto president Onganía, was addressed to CONICET, whose authorities were invited to join the strike of May 30. The note was also sent to the newspapers. On May 29, 1969, very serious disturbances occurred in Córdoba, riots that were later called ``El Cordobazo'' and that marked the beginning of the end of the Onganía government. First the police and then the army intervened to suppress the uprising. Four people were killed, more than 100 were wounded, and many were imprisoned. Meanwhile, the note sent by IAR personnel led to serious disciplinary action by CONICET: Varsavsky was fired as director, and the closure of the IAR was considered. Finally, bureaucratic sanctions were imposed on the staff. Disillusioned by the lack of support from the CIW, Varsavsky resigned from CONICET and gave up scientific research. In 1977, after his nephew Varsavsky was kidnapped and murdered by state terrorism, he left the country. He immigrated with his family to the United States, where he became deputy director of the Institute for Economic Analysis at New York University. He would die in 1983 at the age of 49.

\begin{figure}[!t]
   \includegraphics[width=\columnwidth]{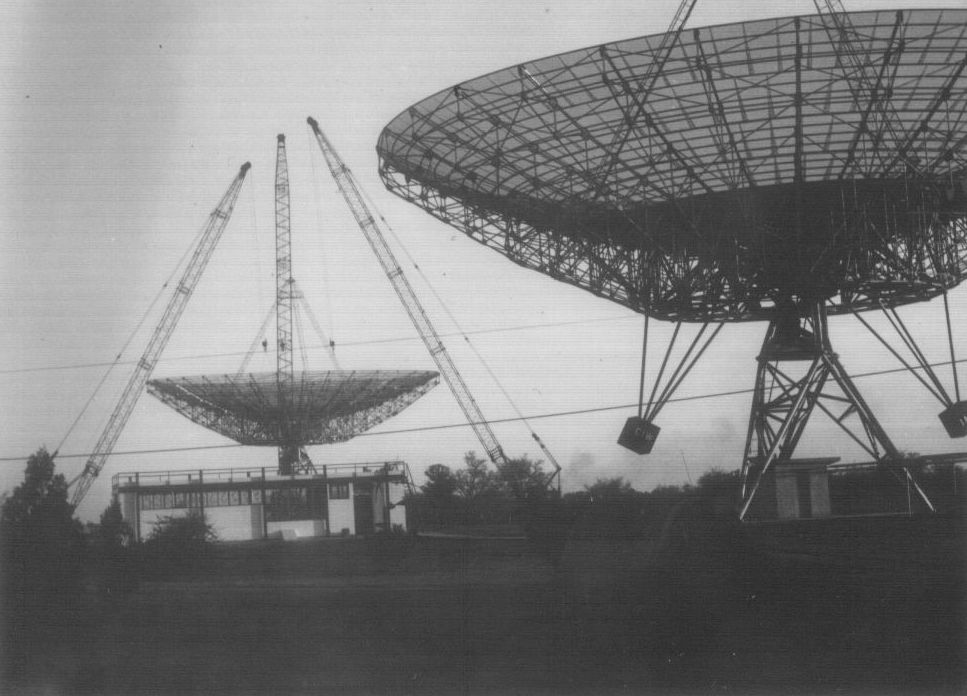}
  \caption{Assembly of the second IAR radio telescope, 1973.}
  \label{fig:a2}
\end{figure}

\begin{figure}[!t]
 \includegraphics[width=\columnwidth]{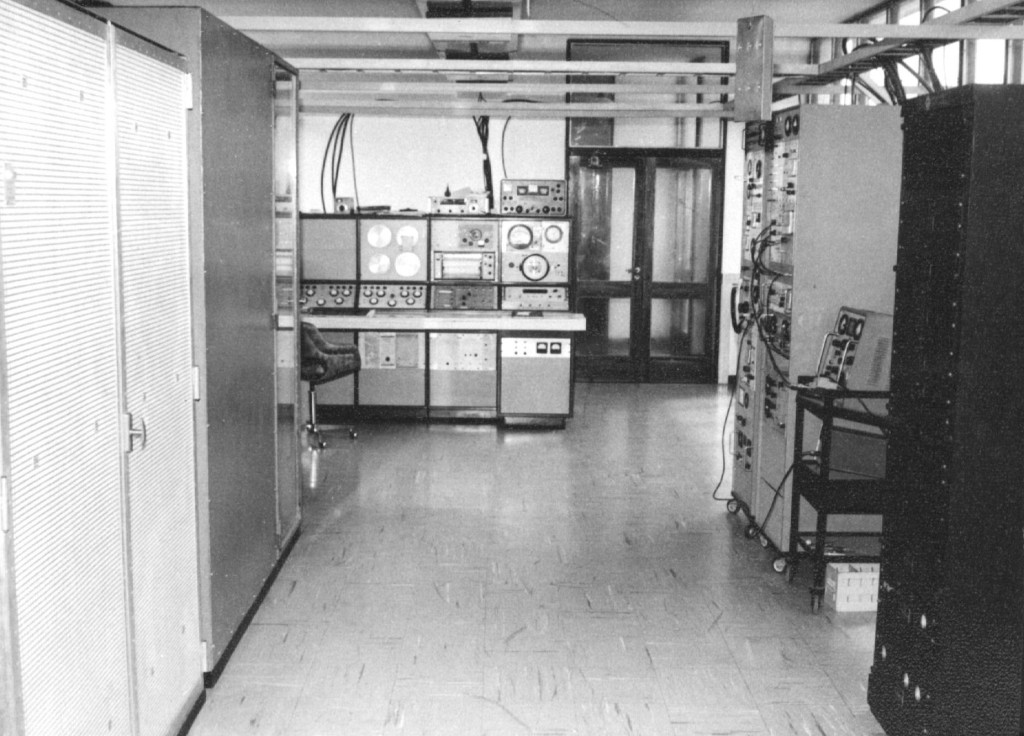}
  \caption{New control room in the building constructed to house it, toward 1976.}
  \label{fig:control}
\end{figure}

Emilio Filloy replaced Varsavsky as interim director until Dr. Kent Turner of DTM was appointed as the new director. During his tenure, there were two important milestones in the history of the IAR: a second, twin radio telescope was built (Fig. \ref{fig:a2}) and a new building housing the telescope control room was inaugurated (Fig. \ref{fig:control}). Turner was also involved in the supervision of students, including Félix Mirabel, who later became one of the most renowned Argentine astronomers. In 1973, elections were held and Héctor Cámpora was elected as the new president. The climate was not favorable for US citizens at that time, so Turner resigned and returned to the US. The new interim director was Raúl Colomb. In 1975, after returning to the country from a long stay in the Netherlands, Esteban Bajaja was appointed director by CONICET.

During Bajaja's tenure as director, a new receiver for the first telescope was built through a renewal agreement with the DTM. This receiver was a significant improvement over the previous one.  It was designed for operation at 21 and 18 cm, with noise temperatures of 83K and 90K, respectively.  A PDP-11/20 computer was used for data acquisition. A new front end was also installed at Antenna 1 on July 21, 1979. It was equipped with a corrugated feeder, carefully designed to attenuate spurious radiation from directions outside the surface of the antenna (the so-called ``spillover'' effect). The filters for the spectral observations consisted of a set of 24 crystal filters, 2.2 kHz wide, 3 kHz apart, for observing very narrow lines of HI or for observing lines of OH. Another set of 112 channels of 10 kHz width and 10 kHz spacing was used for general observations of Galactic HI. Finally, a set of 84 channels of 75.8 kHz (16 kms$^{-1}$), separated by 75.8 kHz, was devoted to the observation of galaxies. All of these filters occupied a large part of the control room. 

With the new equipment, the observatory began a number of important scientific programs, including surveys of HI in galaxies, high-speed clouds, HI in the Magellanic Clouds, a survey of galactic HI at low and intermediate velocities, studies of supernova remnants and HII regions, and observations of recombination lines and OH molecules. The first telescope operated at nearly 100 \% of its capacity. 

After a very successful administration, Bajaja asked for a leave of absence to take a sabbatical year devoted to research. He was replaced as director by Raúl Colomb during the period 1982-1984. Colomb focused on the newest radio telescope, in which a continuum receiver was installed. The idea was to perform a survey of the entire sky in collaboration with the Max Planck Institute for Radio Astronomy (MPIfR) in Bonn. In the northern hemisphere, the observations were made with a 25-meter radio telescope located on Stockert hill in Bad Münstereifel, North Rhine-Westphalia, Germany. A receiver with polarimetric capabilities was developed at the IAR for the project.

Bajaja resumed the directorship in 1984, but soon resigned in discontent because CONICET introduced major changes in the management of scientific budgets. Colomb returned as director for the next 10 years. During these years, most of the efforts were devoted to the surveys. In 1989, an agreement was signed with the Planetary Society to install an 8.4 million channel spectrum analyzer to search for radio signals from extraterrestrial intelligent beings. This instrument, called META II (Mega Channel Extraterrestrial Assay), was built at Harvard by two IAR engineers: Juan Carlos Olalde and Eduardo Hurrell. The instrument became operational on October 12, 1990, and was active for more than two years, accumulating about 9,000 hours of observations. During these observations, 1,600,000 independent spectra of 8.4 million channels each were obtained, more than $10^{13}$ bytes of data. Of the spectra obtained, only those containing peculiar signals were recorded. Finally, 10 suspected cases were selected, but none were confirmed. 

In 1989, the author of this article and Jorge Combi started working at the IAR as telescope operators. In total, they spent more than 300 nights observing with the telescopes during the next two years. This gives an idea of the intensive use of the radio telescopes at that time. In 1992, they both received grants from CONICET to do their Ph.D. research at the IAR, and began to use the instrumental knowledge they had acquired during their years as operators to develop their own research projects on blazars and high-energy particles in the galaxy. Without knowing it, they started relativistic astrophysics in Argentina.

In 1995, Colomb asked for a leave of absence to take a leadership position in the newly created National Commission for Space Activities (CONAE), and was again replaced by Bajaja. Bajaja remained director until September 1997, when he asked for a new sabbatical year and submitted his resignation as director. Ricardo Morras replaced him as interim director. In November 1997, CONICET appointed Marcelo Arnal as the new director, on the recommendation of Bajaja. After only 2 years, Arnal's directorship was terminated by CONICET (see Bajaja 2009), and Ricardo Morras was reappointed as interim director. However, he would remain in charge of the institute for 7 years. This was the time it took for CONICET to call for an opposition to cover the position of director. 

\begin{figure}[!t]
  \includegraphics[width=\columnwidth]{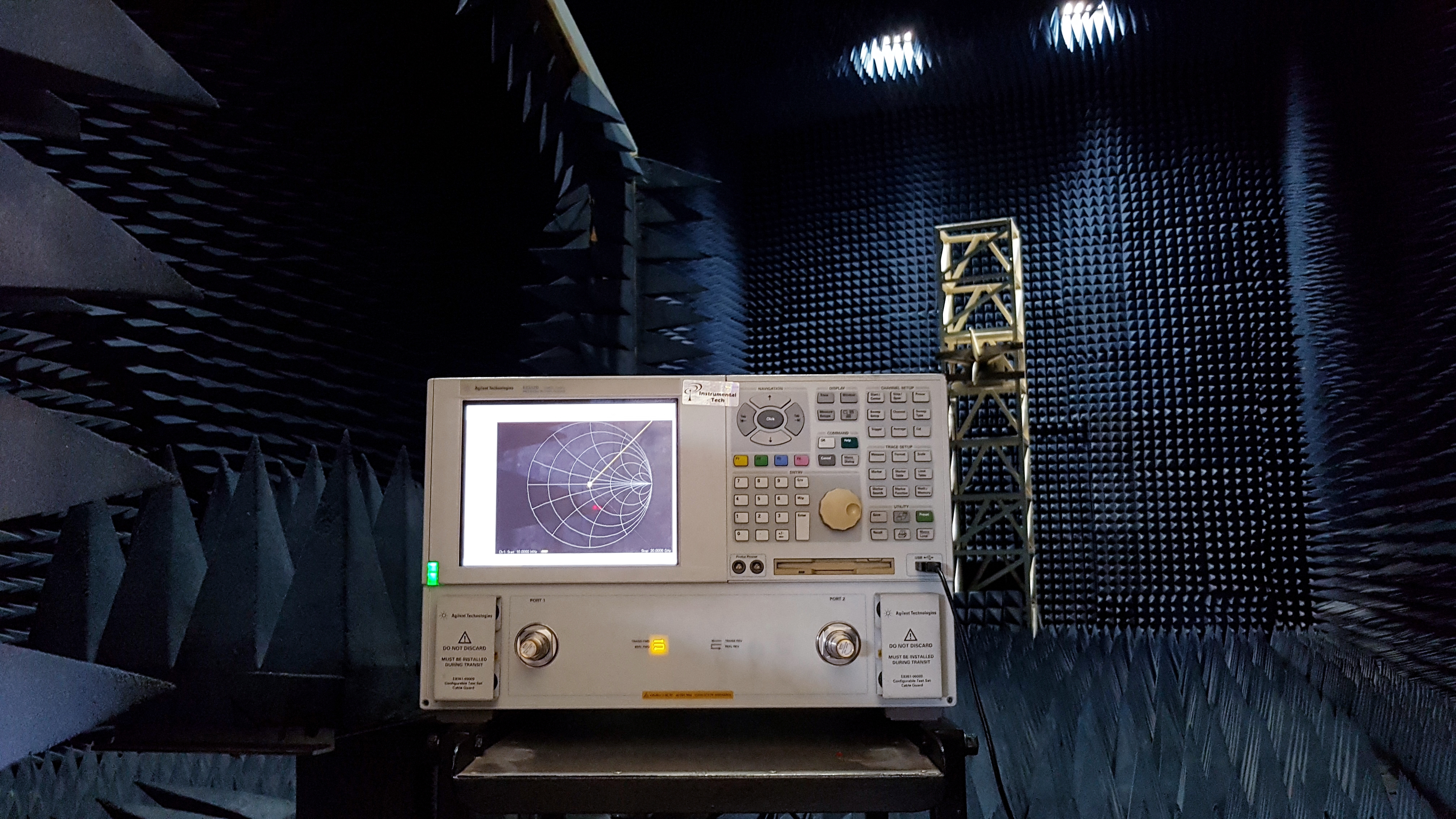}
  \caption{Interior of the anechoic chamber.}.
  \label{fig:chamber}
\end{figure}

\begin{figure}[!t]
  \includegraphics[angle=-90, width=\columnwidth]{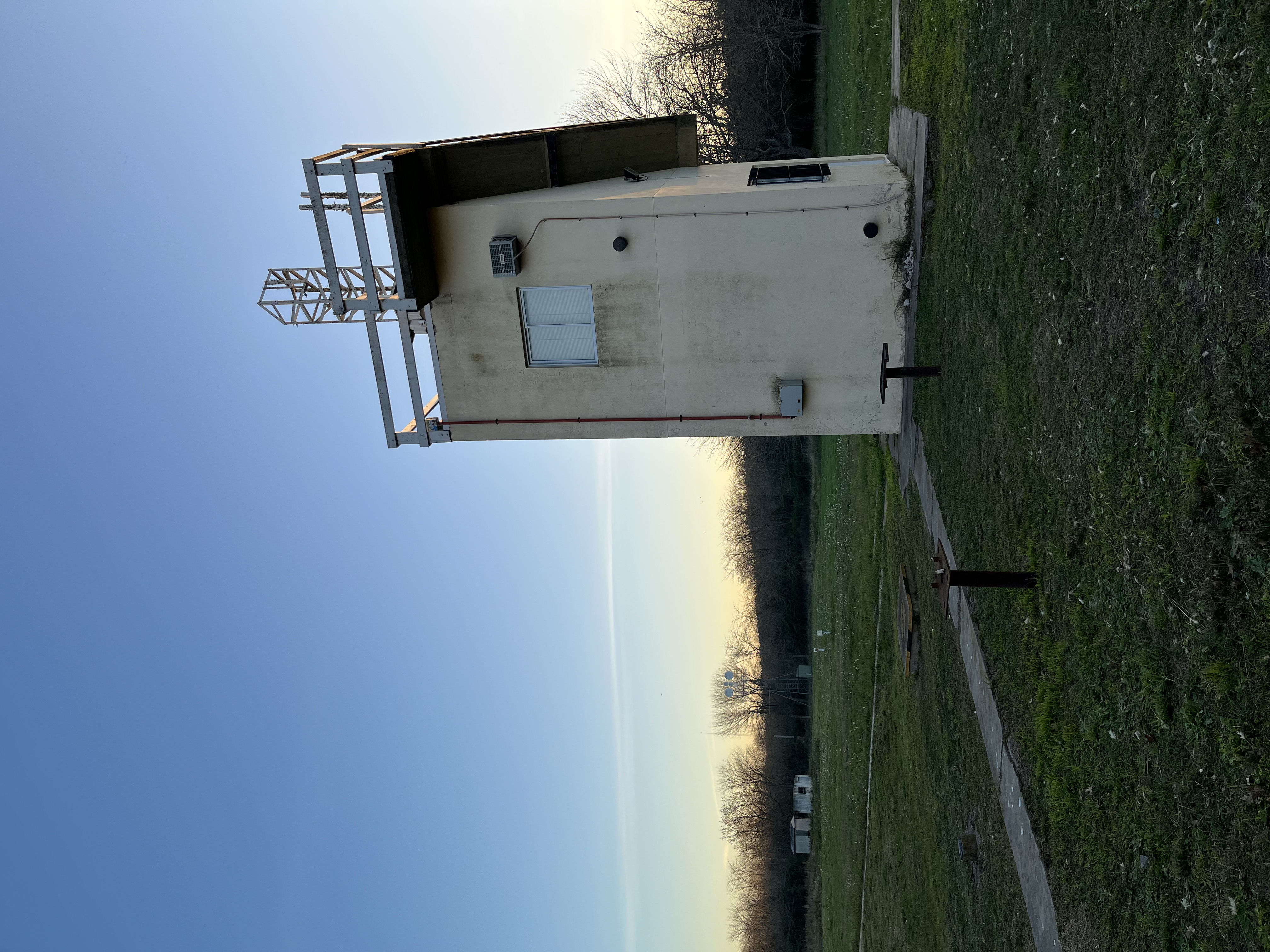}
  \caption{Tower of the far field testing facility.}.
  \label{fig:far-field}
\end{figure}

Morras had to face serious difficulties due to the economic crisis in the country during the years 2001 and 2002 and the consequences of the inherited problems with CONICET. The budget was cut and even the possibility of closing the institute was considered by the authorities; the telescopes had been out of operation during the previous administration and would remain out of service for almost two decades. However, Morras seized the opportunity to begin a fruitful collaboration with CONAE to develop technology for the space program, taking advantage of the strength of the technical staff. As work intensified with new projects related to different satellites, new facilities were built at the Institute. These included a new building for laboratories and a full anechoic chamber (one of only two such facilities in the country, Figure \ref{fig:chamber}), and a far-field test facility (Figure \ref{fig:far-field}). With these new capabilities, the IAR also began to offer technological services. Slowly, under the administration of Morras and the technological supervision of Engs. Juan Sanz and Juan J. Larrarte, the Institute became a recognized technological research and development center within the Argentine system. The number of technical personnel increased dramatically during these years.

From a scientific point of view, most of the research has been done in the field of relativistic astrophysics. In 2000, Romero and Combi, together with Paula Benaglia and Diego Torres, founded the Group of Relativistic Astrophysics and Radio Astronomy (GARRA). This group, led by Romero, was the first to exploit the recent results of gamma-ray instruments, adopting a multi-wavelength approach and interweaving theory and observations. Such an approach, applied systematically for the first time in South America, was extremely successful. Soon the group grew to include up to 20 researchers and students, publishing hundreds of papers and gaining a strong international reputation. Eventually, Romero became professor of relativistic astrophysics at the University of La Plata, Combi created the chair of X-ray astronomy at the same university, Benaglia, who became a leading expert in radio interferometry, founded the group FRINGE, and Torres emigrated first to the USA and then to Spain, where he became director of the Institute of Space Science of Catalonia.    
 
 At the end of 2007, CONICET finally opened the opposition for the directorship and E.M. Arnal returned as director. In a first term he continued the policies of Morras, but after a while he devoted most of his efforts to a new project: a submillimeter radio telescope to be installed in the province of Salta, in the northwestern part of the country, at an altitude of 5000 meters. This project, whose original idea had been conceived long before by Félix Mirabel, was finally approved by the new Ministry of Science (MinCyT) and a substantial budget of several million dollars was allocated. The radio telescope would be purchased by Brazil from the German company Vertex. It was the same design used by ALMA Array in its telescopes. The site and infrastructure would be the responsibility of Argentina, as would the integration of the instrument. The final instrument would be operated as an Argentine-Brazilian facility (see Romero 2020 for a description of the project, called LLAMA). The scale of such a project was underestimated and management problems soon emerged. The consequences for IAR were serious. The director devoted most resources to LLAMA and other projects were affected. Technological activities decreased dramatically and most of the contracted staff moved to other institutions. At the same time, work on the site made little progress. When the telescope finally arrived in Argentina in 2017, Arnal resigned from the project. At the end of the year, he retired as director of the IAR and left the institute. Since he had not appointed a deputy director, the institute was left headless. The situation was very delicate, as the instrument had been left at the international customs, where high fees had to be paid for its storage.  While a new advertisement for the position of Director was organized, two successive interim Directors were appointed by CONICET: Eng. Leandro García and Dr. Leonardo Pellizza. García managed to get the telescope out of customs, but the budget for the project was cut by MinCyT.
   
 \begin{table}[!t]\centering
  \setlength{\tabnotewidth}{\columnwidth}
  \tablecols{2}
  % Stretch the space between table columns 
  \setlength{\tabcolsep}{2.8\tabcolsep}
  \caption{Directors of IAR (1962-2023)} \label{tab:dir}
  \begin{tabular}{lrr}
    \toprule
    Director & \multicolumn{1}{c}{Period} \\
    \midrule
    Carlos M. Varsavsky  & 1962 -- 1969  \\
    Emilio Fillioy & 1969 -- 1971  \\
    Kent Turner & 1971 -- 1973  \\
    Fernando Raúl Colomb & 1973 -- 1975 \\
    Esteban Bajaja & 1975 --  1982  \\
    Fernando Raúl Colomb & 1982 -- 1984  \\
    Esteban Bajaja & 1984 -- 1985  \\
    Fernando Raúl Colomb & 1985 -- 1995  \\
    Esteban Bajaja & 1995 - 1997  \\
    Edmundo M. Arnal & 1997 -- 1999  \\
    Ricardo Morras & 1999 -- 2007  \\
    Edmundo M. Arnal & 2007 -- 2017  \\
    Leandro García & 2018 -- 2018    \\
    Leonardo Pellizza & 2018 -- 2018   \\
    Gustavo E. Romero & 2018 -- 20...  \\
    \bottomrule
   % \tabnotetext{a}{Note the use of \CS{lowercase} to prevent the $ from being converted to upper case.}
  \end{tabular}
\end{table}

\begin{figure}[!t]
  \includegraphics[width=\columnwidth]{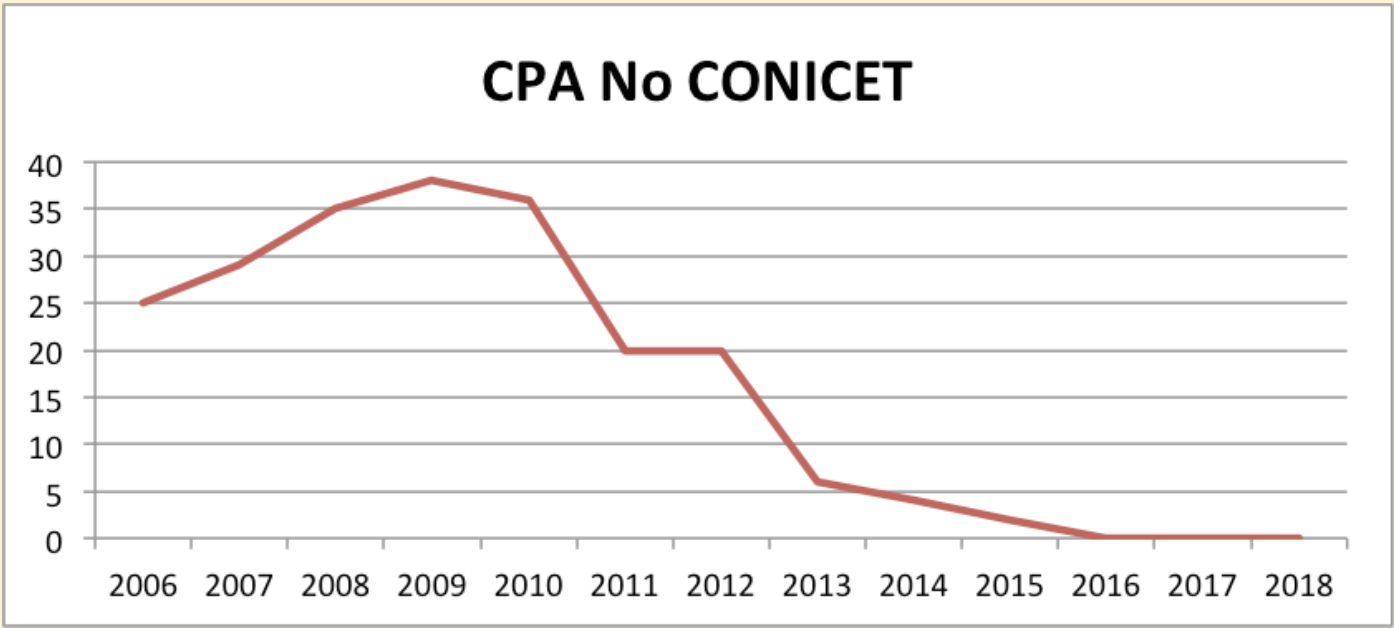}
  \caption{Number of contracted technical personnel through the years.}.
  \label{fig:staff}
\end{figure}

In August 2018, Gustavo E. Romero took over as the new director of the IAR. The situation was difficult: the LLAMA project was stopped by MinCyT, there was no other institutional project, the deterioration of the Institute's facilities was unprecedented, the IAR's bank accounts were frozen by court order, the Institute was completely isolated within the Argentine science and technology system (there was no communication even with the nearby AGGO Observatory), the radio telescopes had been out of service for about 20 years, technological development and the associated technology transfer had stopped, the technical staff was decimated: about 40 professionals left the IAR in the previous 10 years (see Fig. \ref{fig:staff}). 

MinCyT soon decided to transfer the management of LLAMA to ITEDA, an institute dedicated to astroparticle detection and related technologies. After a year and a half, when the government changed in 2019, the responsibility for integrating the instrument was transferred again, this time to INVAP, a government technology company. IAR remained as a consultant and contractor for some specific tasks. Romero was then able to devote his full attention to the reorganization of the Institute. After resolving the legal problems left by LLAMA's management, he implemented a new structure to allow for better coordination and cooperation in achieving institutional goals. Two large areas were created: scientific and technological, each with a person in charge. The technological area has two large sectors, the Observatory and the Technology Development and Transfer Sector, again each with a head appointed by the director. Under these structures, several departments have been created, each with its own supervisor: the Electronic Department, the Mechanical Department, the Systems Department and the Outreach Department. All these departments have their own structure and must provide the services necessary to achieve the institutional objectives. This system proved to be very efficient.
 
The next step was to obtain funding for a new institutional project. To achieve this, a new security plan was proposed to CONICET: the six hectares of the IAR property were fenced off, access to the institute was automated, and the entire property was monitored with more than 40 cameras. Security was unified with that of the neighboring AGGO observatory, with which an agreement was signed. With half of the budget spent on security guards in the past, this new system, fully developed and implemented by IAR staff in record time, freed up significant funds for other purposes.

\begin{figure}[!t]
  \includegraphics[width=\columnwidth]{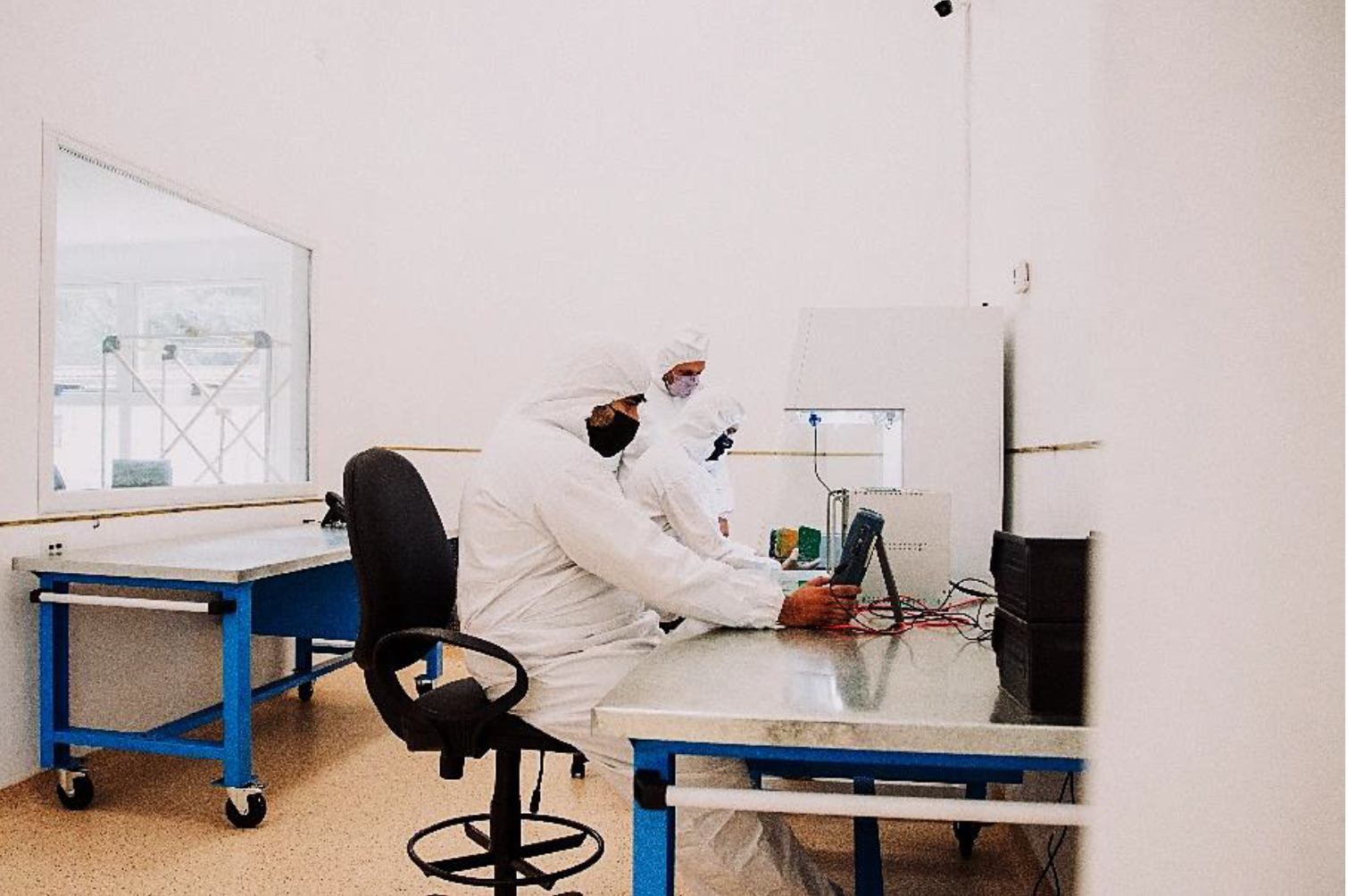}
  \caption{New clear laboratory at IAR (inaugurated in 2021).}
  \label{fig:clean}
\end{figure}

These purposes were dictated by the new institutional project: the recovery of the radio telescopes, the complete upgrade of their technology, and the development of a new astronomical facility: the so-called Multipurpose Interferometric Array (MIA), which is now progressing with the construction of a demonstrator. Technology transfer has also been strongly promoted and new projects have emerged: from nanosatellites to medical equipment to combat the COVID-19 pandemic. Part of the funds that had been obtained from the services were used for the improvement of the working conditions. The progress stimulated renewed support from CONICET, CIC and UNLP. New facilities were built, including a clean laboratory (Fig. \ref{fig:clean}). International support was strong, with generous donations and cooperation from the Rochester Institute of Technology (RIT), the Harvard-Smithsonian Observatory, the Collaboration for Astronomy Signal Processing and Electronics Research (CASPER), the Parkes Observatory, and others. More than 15 different types of agreements have been signed in 4 years. Also, the association with CONAE was strengthened with a new agreement and digital receivers for radio astronomy were developed for the two deep space antennas operating in Argentina. The first one was installed in December 2022.

\begin{figure}[!t]
  \includegraphics[width=\columnwidth]{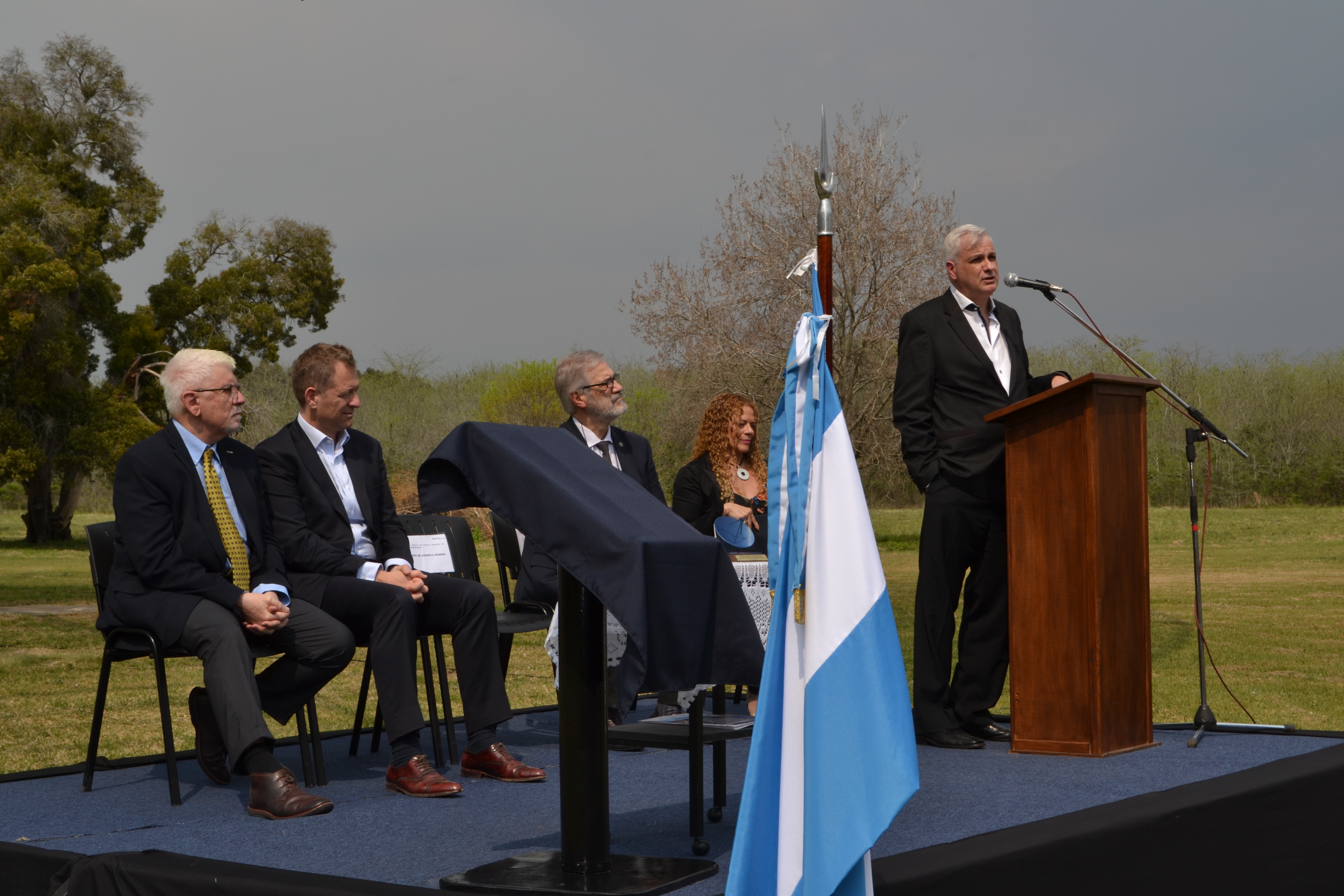}
  \caption{Gustavo E. Romero, with authorities from CONICET, CIC, and MinCyT, during the baptism of the fully restored and upgraded telescopes.}
  \label{fig:Bautismo}
\end{figure}

\begin{figure}[!t]
  \includegraphics[width=\columnwidth]{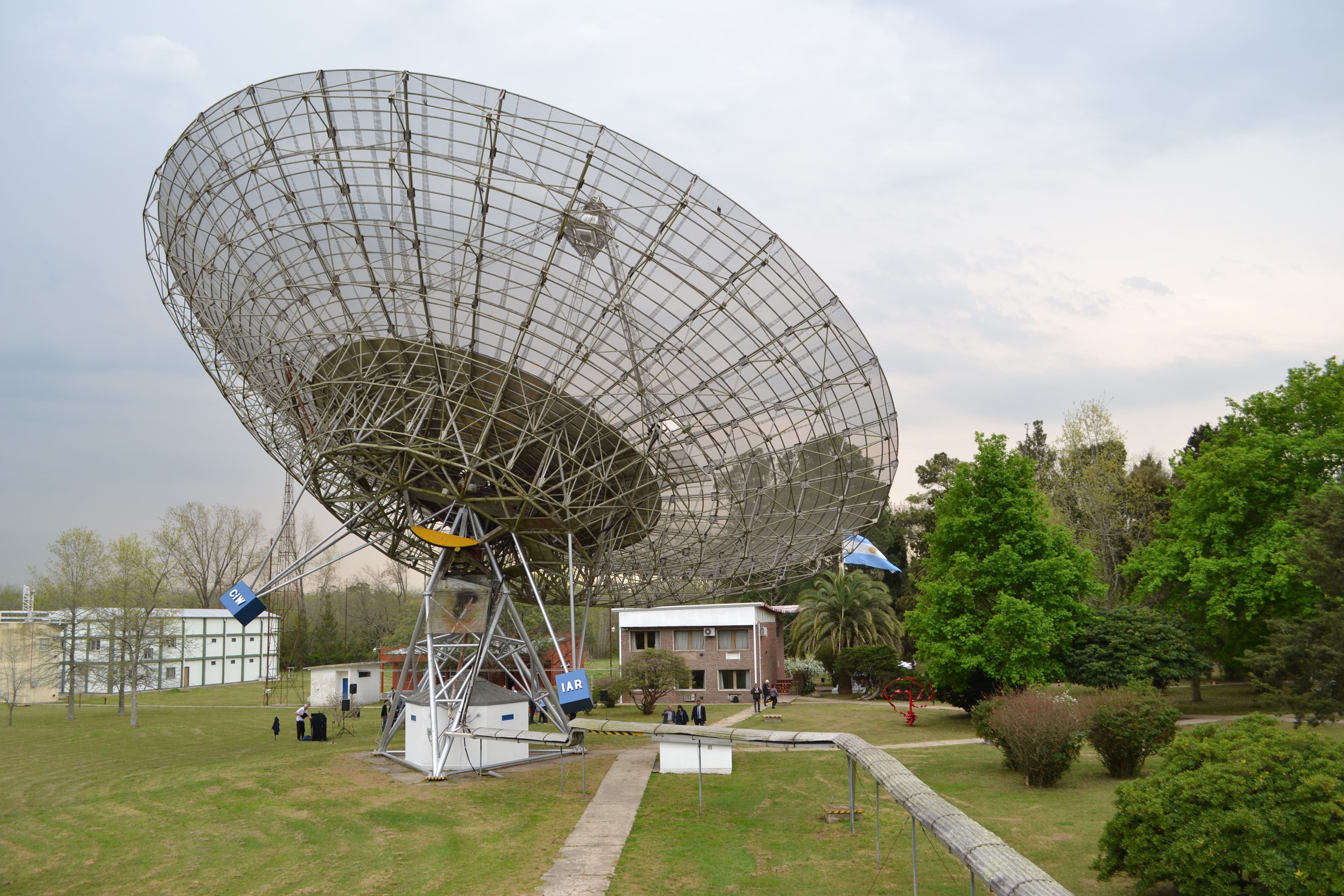}
  \caption{IAR in 2019, the day the radio telescopes were baptise.}
  \label{fig:Bautismo2}
\end{figure}

In 2019, both radio telescopes are IAR were baptized and the observatory was reopened (Figs. \ref{fig:Bautismo} and \ref{fig:Bautismo2}). The telescopes, which had been called ``Antenna 1'' and ``Antenna 2'' for almost 50 years, were now named after Carlos M. Varsavsky and Esteban Bajaja, respectively. By the time of its 60th anniversary, the IAR was once again a leading research institute in Argentina.

\section{Some highlights}
\label{sec:highlights}

\begin{figure}[!t]
  \includegraphics[width=\columnwidth]{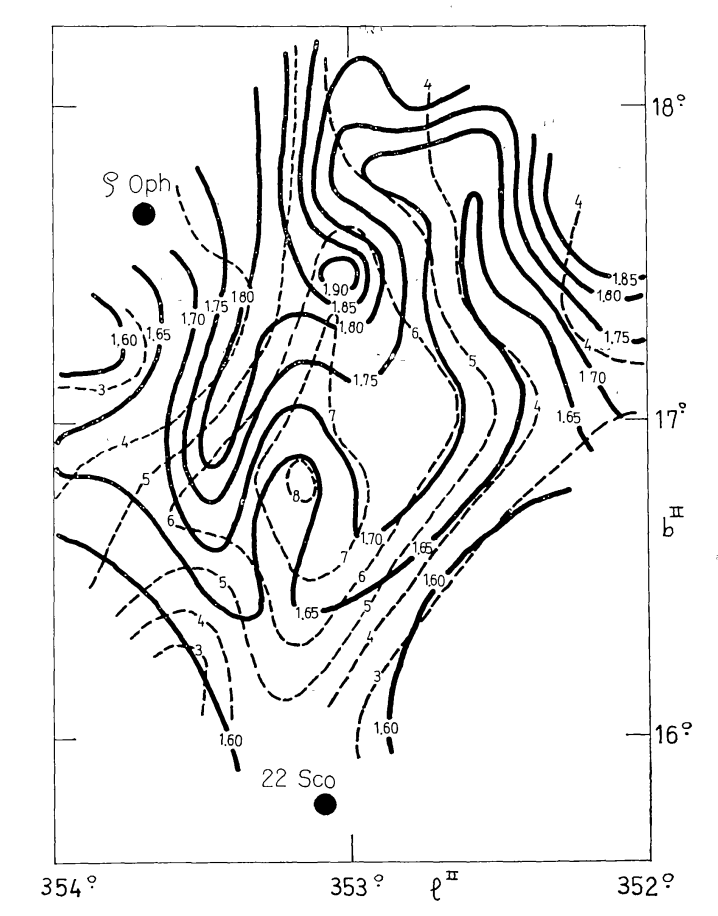}
  \caption{First HI map made with data obtained with IAR first radio telescope (M\'esz\'aros 1968). }
  \label{fig:Meszaros1968}
\end{figure}

The first scientific paper published  with affiliation at ``IAR'' was a review by Varsavsky (1966) on the detectability of molecular hydrogen in interstellar space. This was followed by several papers based on observations made with other observatories. The first paper based entirely on observations made with the IAR radio telescope was M\'esz\'aros (1968). This paper presented the first HI map constructed from data obtained with the new instrument. It was an observation of a dark cloud near $\rho$ Ophiuchi (see Fig. \ref{fig:Meszaros1968}). Towards the end of the 1960s, a steady stream of publications from the IAR began. So far, more than 1300 papers have been published with the affiliation ``Instituto Argentino de Radioastronomía''. Behind these papers is a huge amount of research. There are three periods in which the activity increased significantly: first, between 1980 and 1990, with the installation of the new receiver in the first radio telescope and the corresponding correlator. Then, between 2003 and 2013, by a burst of activity in theory by the GARRA group. And finally, the current period with many contributions from all groups, FRINGE, GARRA and PuMA. 

\begin{figure}[!t]
  \includegraphics[width=\columnwidth]{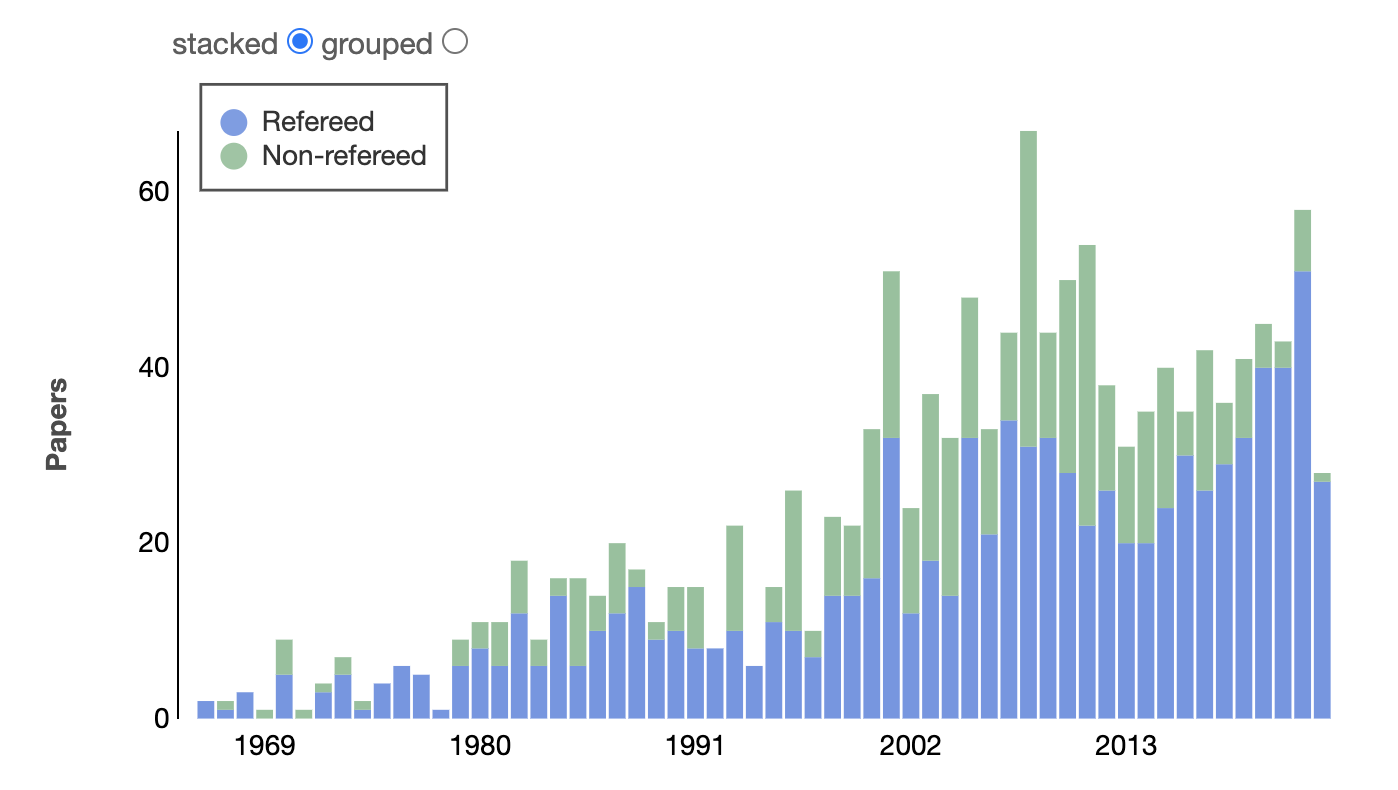}
  \caption{Papers published by authors from IAR through the history of the institute. Altogether, more than 1300 papers have been published so far. Source: NASA's ADS. }
  \label{fig:papers}
\end{figure}

\begin{figure}[!t]
  \includegraphics[width=\columnwidth]{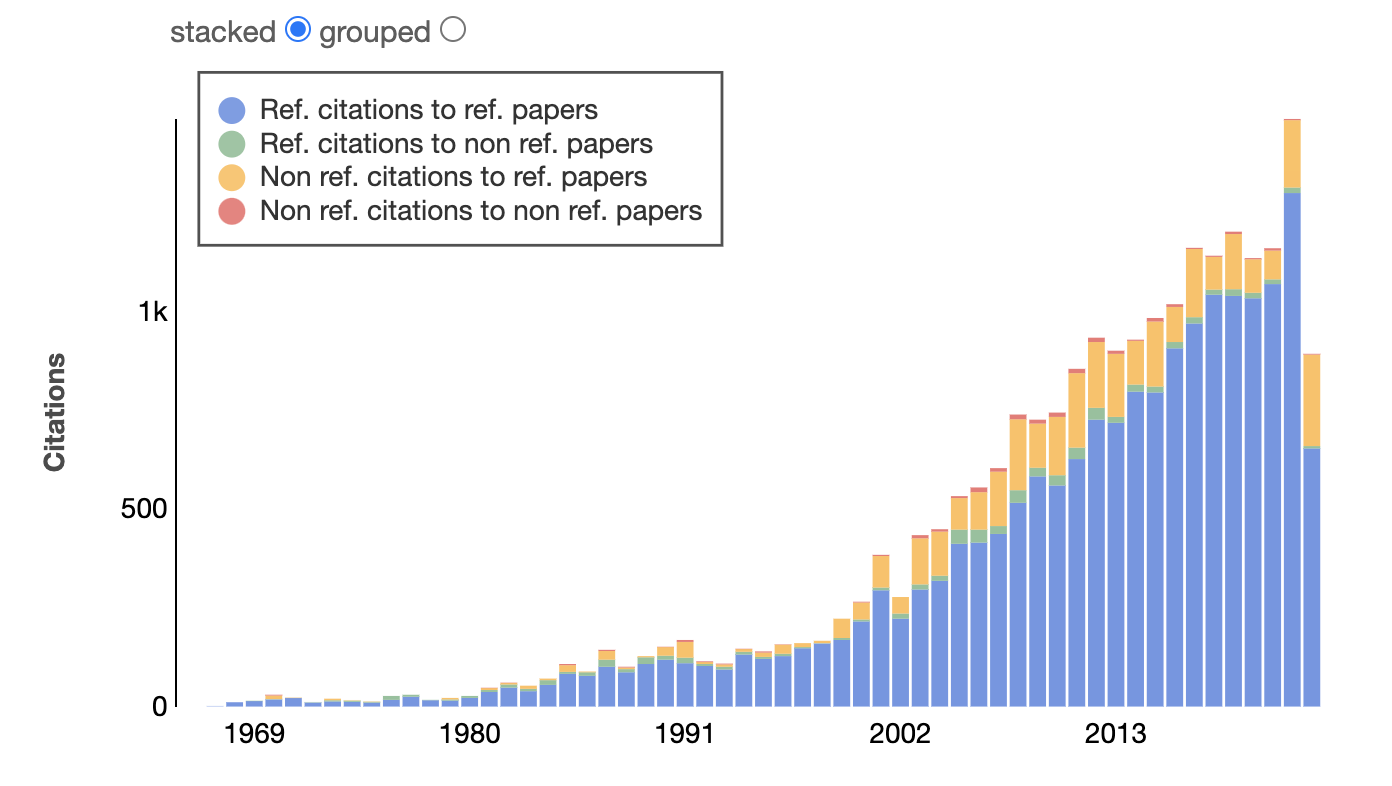}
  \caption{Number of citations of IAR's publications through the years. Source: NASA's ADS.}
  \label{fig:citations}
\end{figure}

In Fig. \ref{fig:papers} I show the evolution of the number of papers published per year by IAR. The absolute maximum, regarding articles in refereed journals, occurred in 2022, with 51 publications. Figure \ref{fig:citations} shows the number of citations per year. The citations have been growing steadily, exceeding 1000 per year in the last 5 years. In the remainder of this section, I will mention a few highlights among the published research.

\begin{figure}[!t]
  \includegraphics[width=\columnwidth]{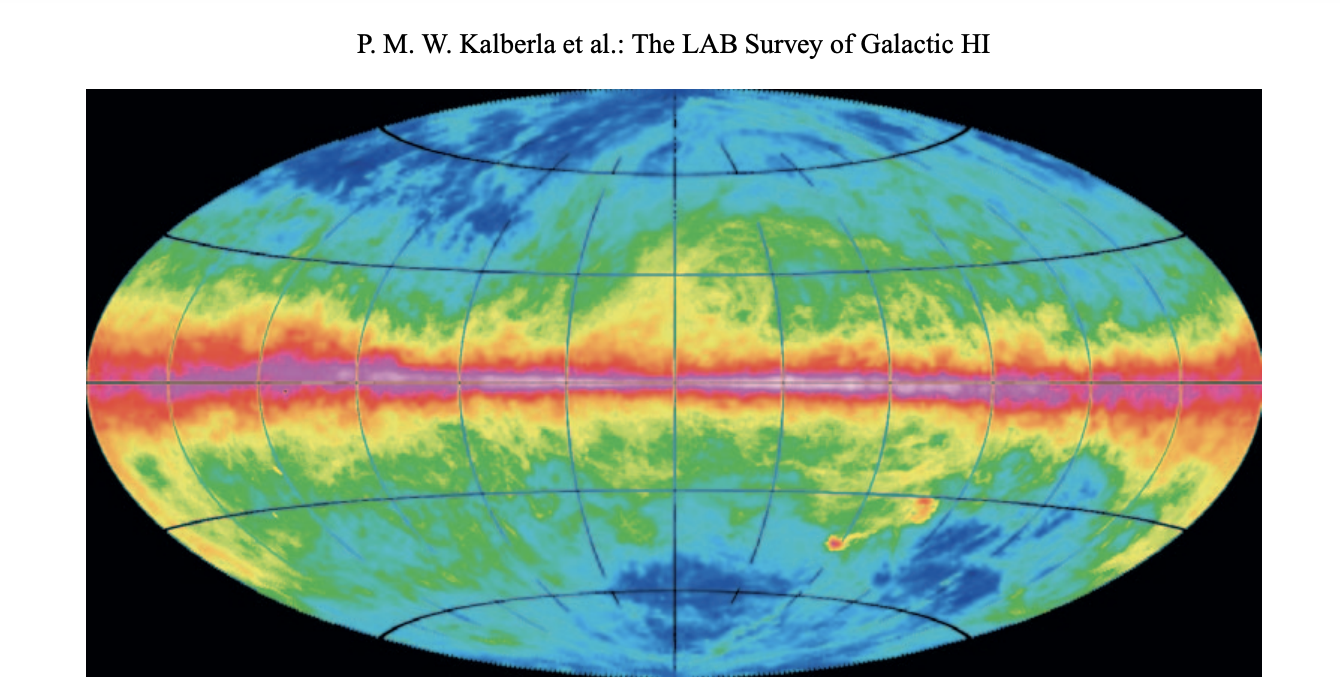}
  \caption{HI emission integrated over the velocity range $-400 < v < +400$ km s$^{-1}$ in the LAB dataset, shown in an Aitoff projection (Kalberla et al. 2005). }
  \label{fig:HI}
\end{figure}

In 1976 the first results of a survey of HI were published by Colomb et al. (1976). It was not until 2005, 30 years later, that the final results of the Leiden/Argentine/Bonn (LAB) Survey of Galactic HI were published (Kalberla et al. 2005). This database has become extremely useful to the astrophysical community, as evidenced by the large number of citations (3200+). Figure \ref{fig:HI} shows the HI map of the whole sky in the velocity range $-400 < v < +400$ km s$^{-1}$.

The continuum survey was published in 2001 (Testori et al. 2001, Reich et al. 2001) and the associated linear polarization survey in 2008 (Testori et al. 2008). The radio telescopes have also been used to study individual sources, both in the HI line and in the continuum. Mirabel and Turner (1975) made observations of high-speed clouds. This line of research was continued for many years by Mirabel and Morras. Colomb \& Dubner (1980) published a first paper on supernova remnants (SNR) based on an analysis of the atomic gas.  Combi et al. (1999) detected the Vela Jr. SNR in radio, using the second IAR radio telescope, shortly after its discovery in X-rays. Romero et al. (1994) reported the detection of what would later be called an ``extreme scattering event'' in the direction of a compact active galactic nucleus. Combi \& Romero (1995) carried out an investigation of the possible radio counterparts of an unidentified gamma-ray source. This paper also contained the first theoretical study of high-energy astrophysics made in the country. It would initiate a far-reaching research program that is still ongoing.

The first paper entirely devoted to theory was published by Olano (1982), with a simple but very successful model of the local gas associated with Gould's belt. Romero et al. (1995a) presented the first of many papers on the physics of relativistic jets, while Romero et al. (1995b) started research on gravitational lensing at the IAR. Such a research program continues to this day, now including studies using numerical relativity. The theoretical study of microquasars, originally discovered by Mirabel working in France in the early 1990s, started with Kaufman Bernadó et al. (2002) and Romero et al. (2002). This research program, which would lead to several Ph.D. theses, continues, as do numerical studies related to this topic (Romero et al. 2007). Benaglia \& Romero (2003) predicted that colliding wind binaries should produce gamma rays and presented some early estimates. This work was followed by many others, both theoretical and observational, on massive stars, their winds, and their interactions.

  Important discoveries have also been made by observations with other instruments. Brinks \& Bajaja (1986) found a large number of holes in the HI distribution of the Andromeda galaxy using the Westerbork radio telescope. Benaglia et al. (2010) detected the first non-thermal bowshock around a runaway star. This was followed by a catalog of bowshocks (Peri et al. 2012, 2015) and many theoretical developments (e.g. del Valle \& Romero 2012).

\section{Present}
\label{sec:present}

\begin{figure}[!t]
  \includegraphics[width=\columnwidth]{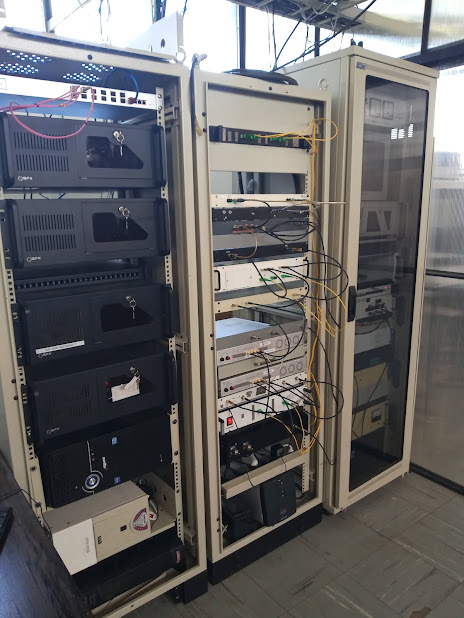}
  \caption{Digital back-end based on ROACH boards developed by CASPER collaboration and storage in the control room of IAR's radio telescopes.}
  \label{fig:back-end}
\end{figure}

\begin{figure}[!t]
  \includegraphics[width=\columnwidth]{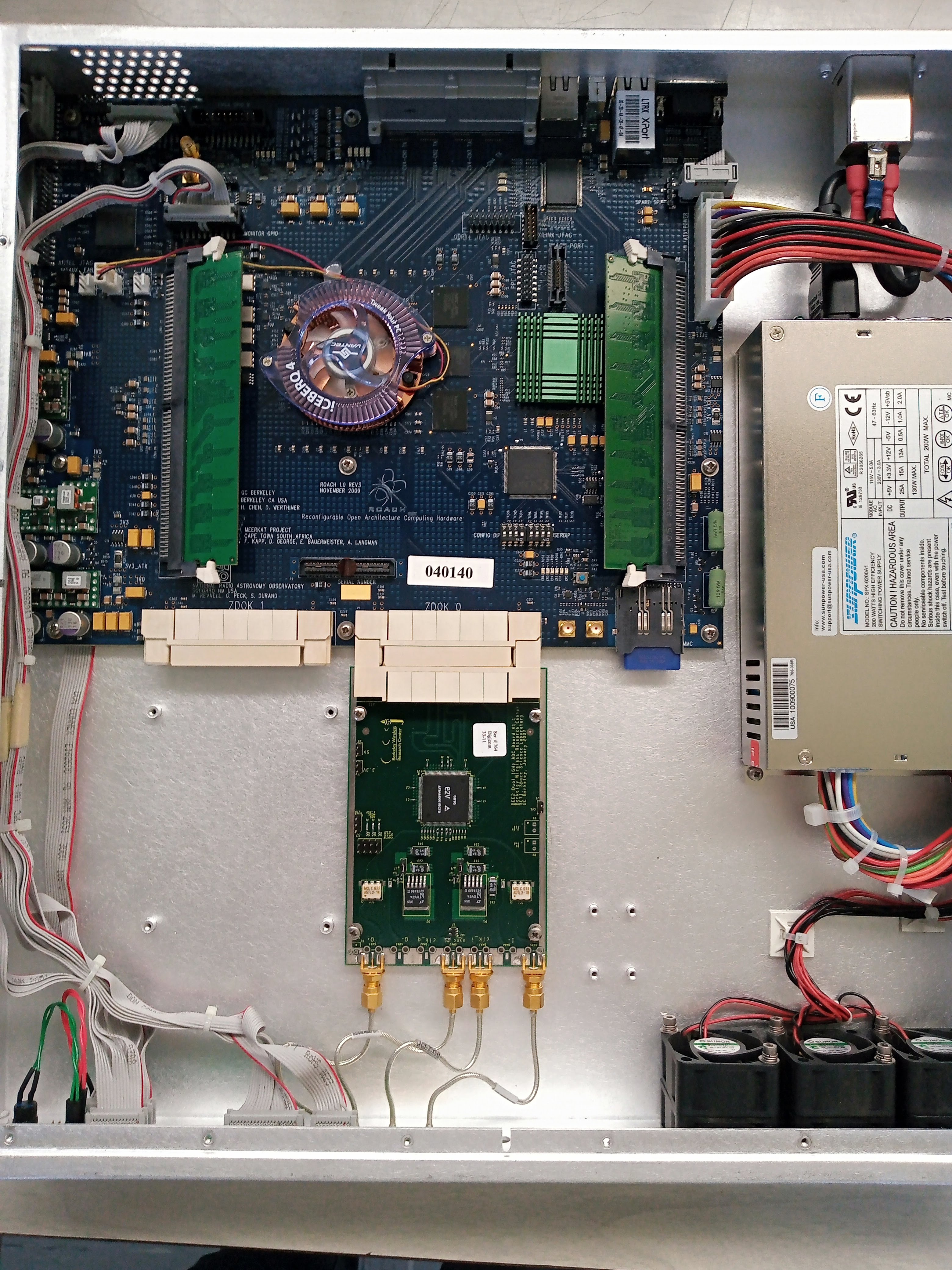}
  \caption{View of a ROACH board integrated to the back-end.}
  \label{fig:roach}
\end{figure}

  Today, both radio telescopes have the capability to operate remotely and continuously to study different astronomical objects. After various upgrades (see Gancio et al. 2020), they are mainly dedicated to the observation of fast variability in different types of sources such as neutron stars, magnetars, blazars and pulsar timing, which will allow the detection and study of gravitational waves (see https://nanograv.org/). The telescopes have digital receivers using state-of-the-art hardware platforms based on high-performance FPGAs developed by the CASPER (Collaboration for Astronomy Signal Processing and Electronics Research) group. With these instruments and new front ends, the telescopes have a bandwidth of 400 MHz each, with two polarizations. The Varsavsky telescope has a center frequency of 1420 MHz, while the Bajaja telescope is centered at 1300 MHz. The temperature of both receivers is less than 80 K. The pulsar observations are performed with 512 channels and a band of 400 MHz, a typical bandwidth of each channel of 781 kHz, and an integration time of 73 microseconds. The back-end is shown in Fig. \ref{fig:back-end}. Fig. \ref{fig:roach} shows a detail of a ROACH board.
  
The time reference is provided by GPS, GNSS, and an atomic clock. All connections are made with optical fibre.  Operation is fully automatized.

\begin{figure}[!t]
  \includegraphics[width=\columnwidth]{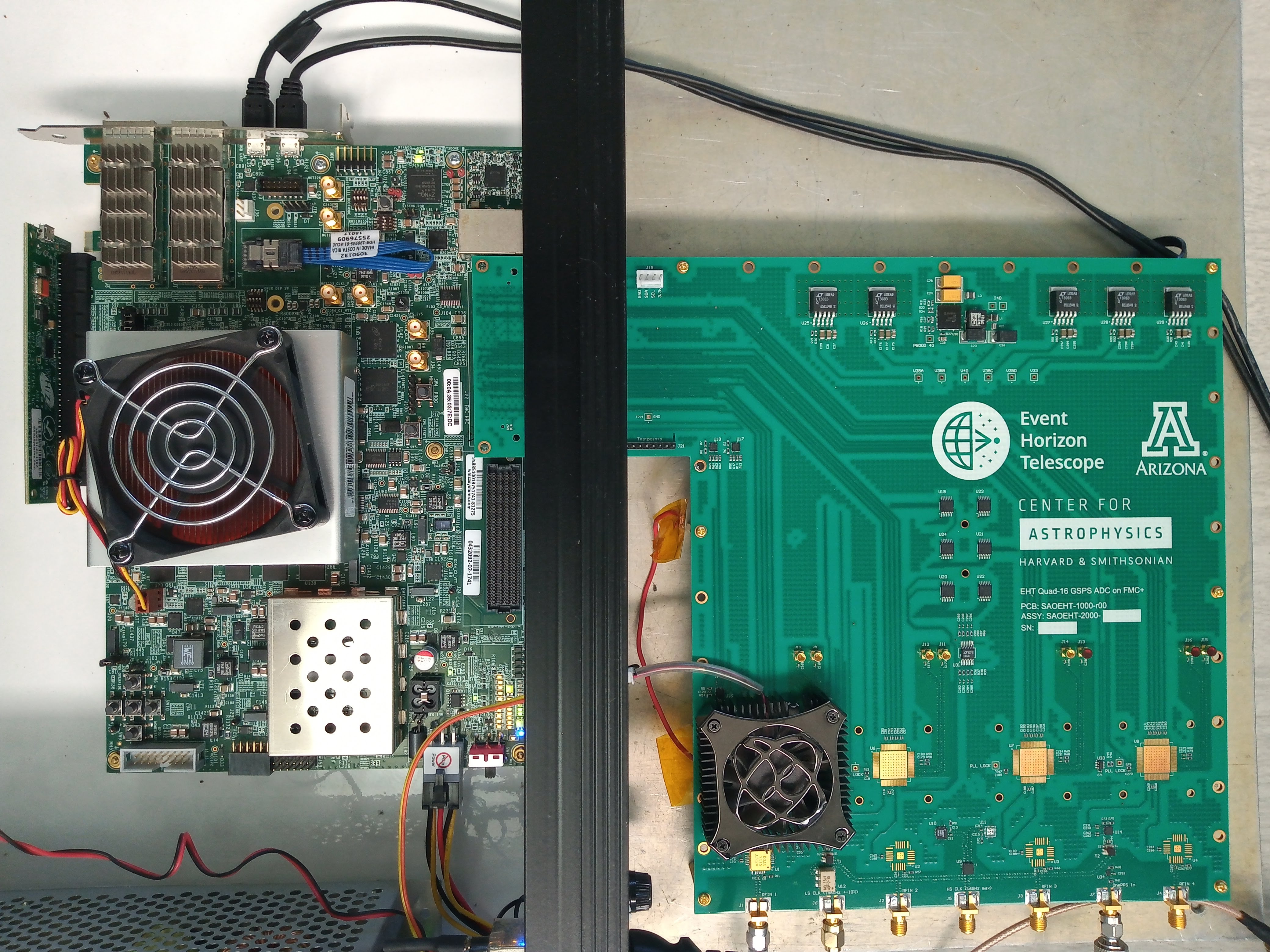}
  \caption{View of the prototype digital receiver for the Next Generation Event Horizon Telescope, being developed at IAR in collaboration with the Harvard-Smithsonian Astronomical Observatory.}
  \label{fig:ngEHT}
\end{figure}

In addition to hardware of its own observatory, the IAR contributes with developments for other institutions such as CONAE and the Next Generation Event Horizon Telescope (Fig \ref{fig:ngEHT}). 

\begin{figure}[!t]
  \includegraphics[width=\columnwidth]{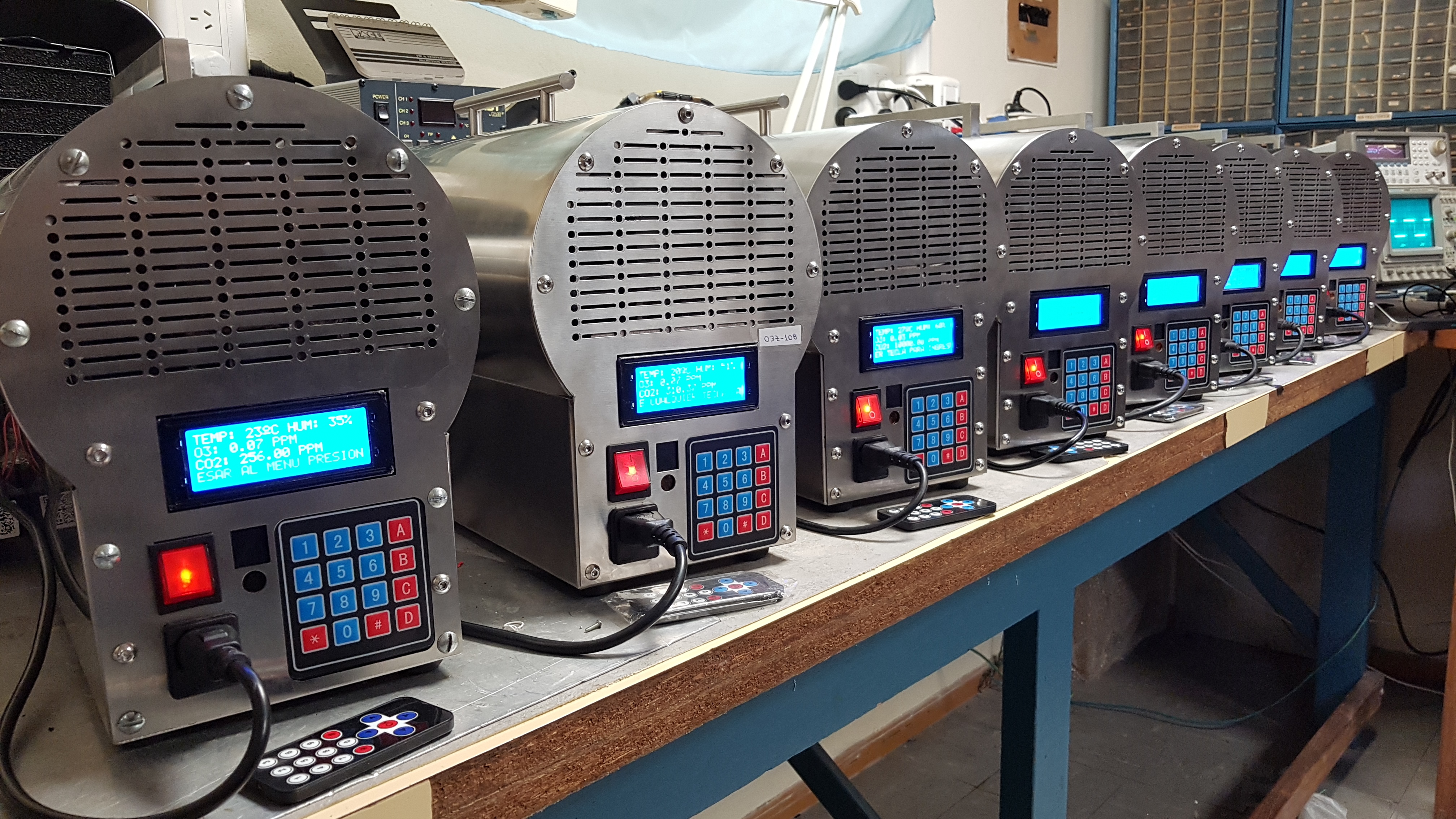}
  \caption{Assembly line of ozone reactors for sterilization close spaces.}
  \label{fig:ozone}
\end{figure}

\begin{figure}[!t]
  \includegraphics[width=\columnwidth]{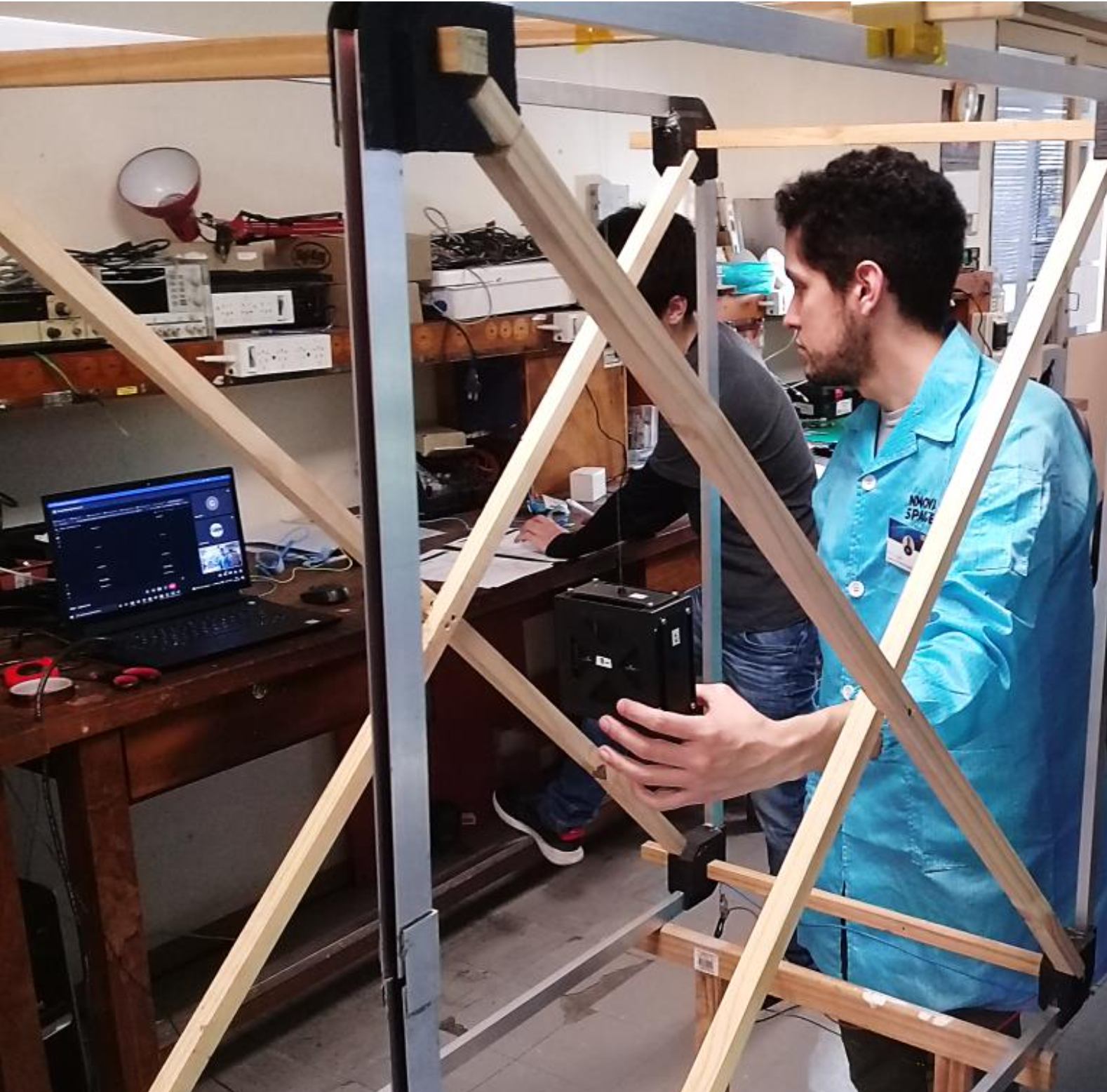}
  \caption{PocketQube Helmholtz cage for testing satellite magnetotorquers.}
  \label{fig:nano}
\end{figure}

\begin{figure}[!t]
  \includegraphics[width=\columnwidth]{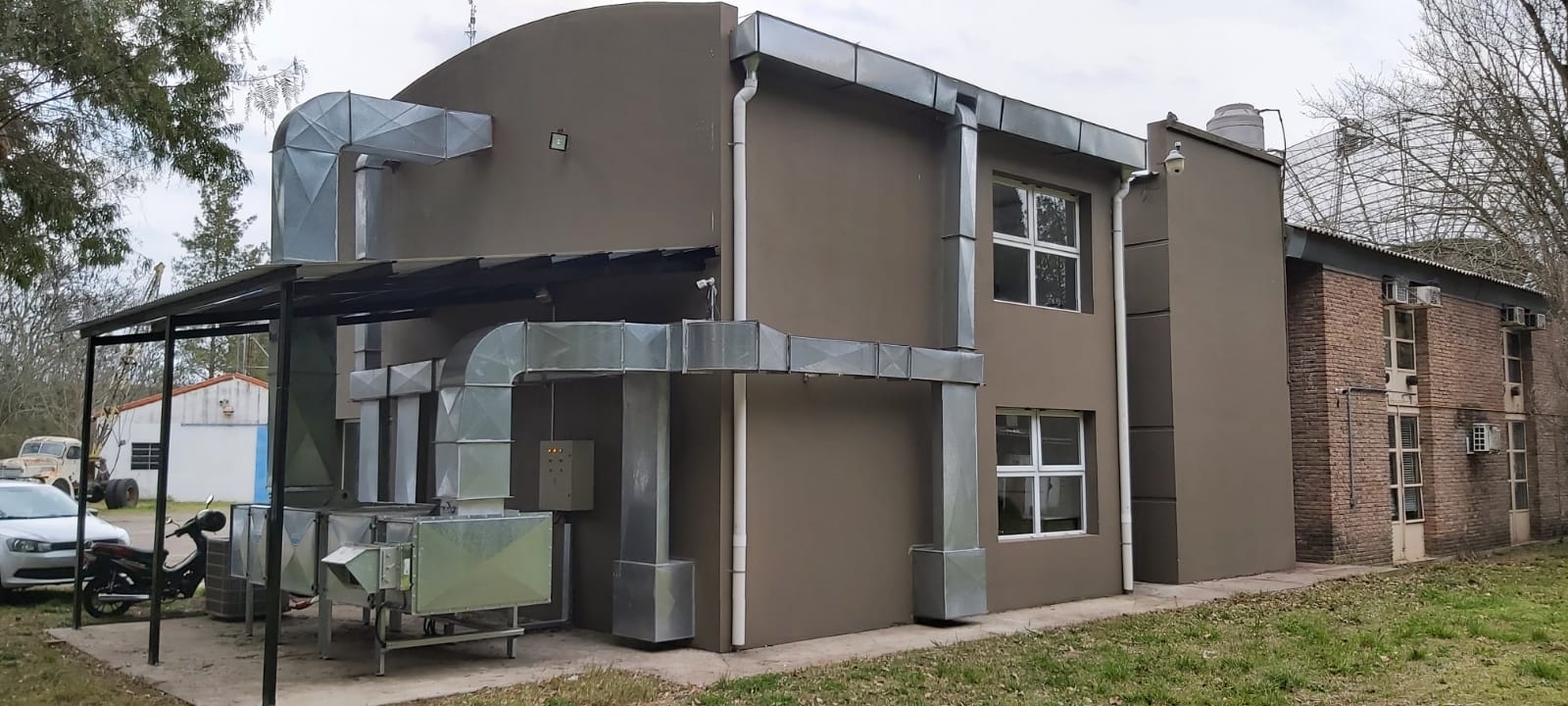}
  \caption{New building with clean labs.}
  \label{fig:clear-exterior}
\end{figure}

\begin{figure}[!t]
  \includegraphics[width=\columnwidth]{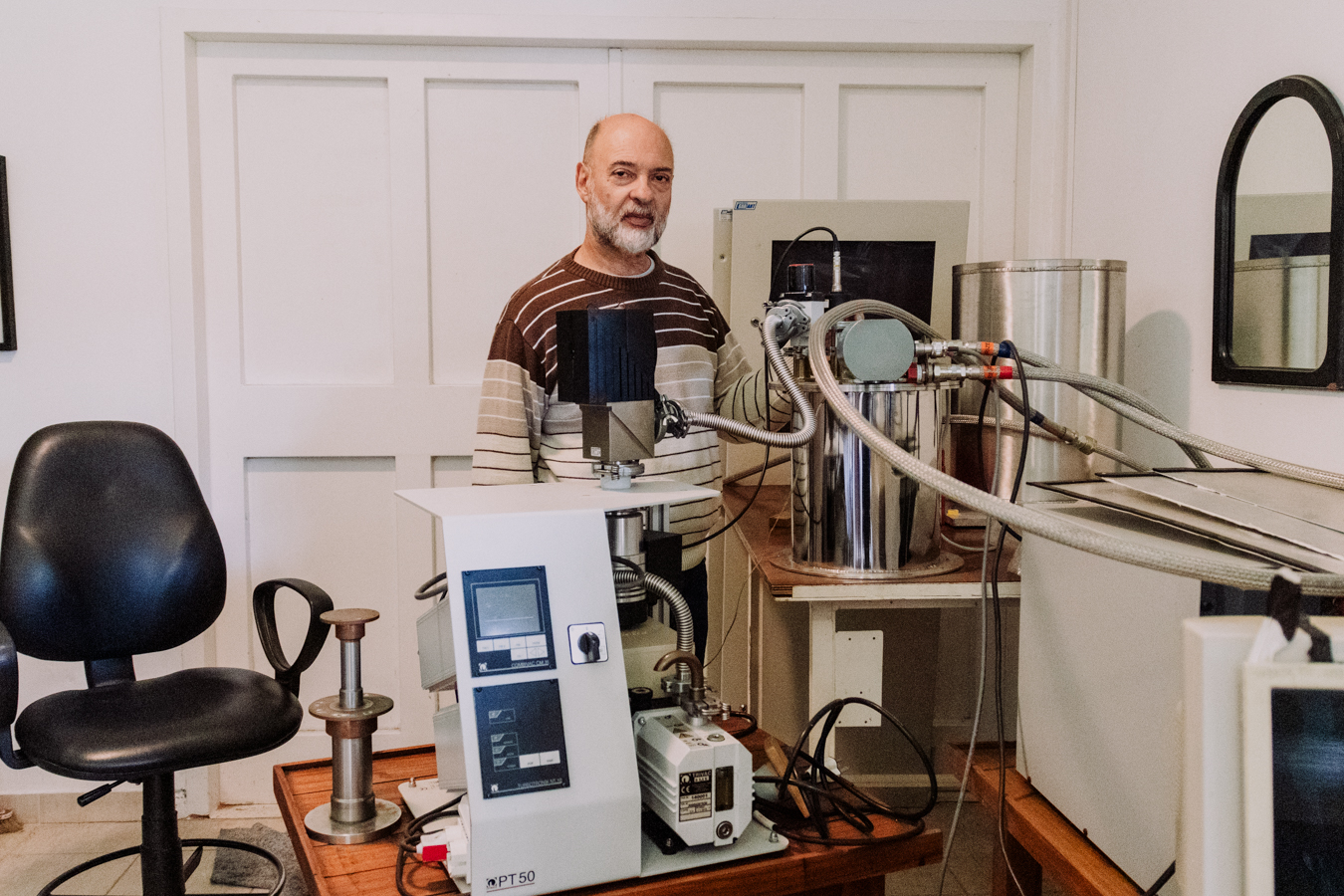}
  \caption{Cryogenic laboratory.}
  \label{fig:cryogenic}
\end{figure}

\begin{figure*}
    \centering
    \includegraphics[width=0.95\textwidth]{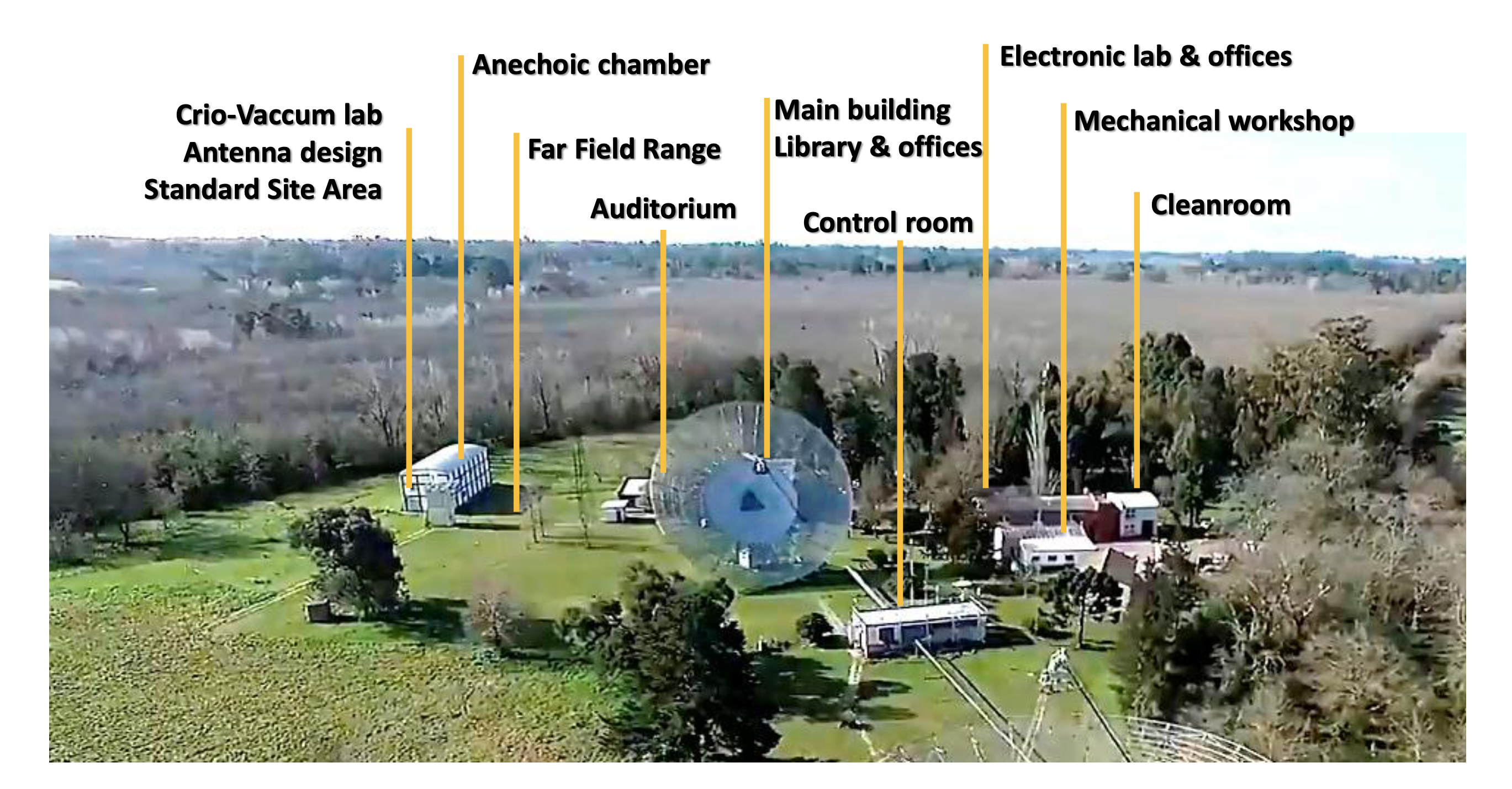}
    \caption{Aerial view of the central part of the IAR, 2022.}
    \label{fig:view}
\end{figure*}

In addition to radio astronomical work and theoretical research, the IAR today is also dedicated to the development of technological innovations and applications for general use. These range from ozone reactors (Fig. \ref{fig:ozone}) to nanosatellites (Fig. \ref{fig:nano}). New facilities have been added for such tasks (Figs. \ref{fig:clear-exterior} and \ref{fig:cryogenic}). The new ISO 8 cleanroom is class 100,000 with a class 100 horizontal laminar flow cabinet. The main cleanroom is 32 m$^2$ and there is an auxiliary storage room. A view of most of the facilities available at IAR today is shown in Fig. \ref{fig:view}.

%\begin{figure}[!t]
  %\includegraphics[width=\columnwidth]{view}
  %\caption{Aerial view of the central part of the IAR, 2022.}
  %\label{fig:view}
%\end{figure}

\section{A view of the future}
\label{sec:future}

The two 30-meter radio telescopes at the IAR are now equipped with state-of-the-art electronics. However, a challenge for the immediate future is to dramatically increase the storage capacity. Both pulsar timing observations and searches for fast radio bursts generate huge amounts of data. The current administration of the IAR is well aware of this problem and it is expected that a new acquisition system with up to 1 PB of storage will be implemented soon.

\begin{figure}[!t]
  \includegraphics[width=\columnwidth]{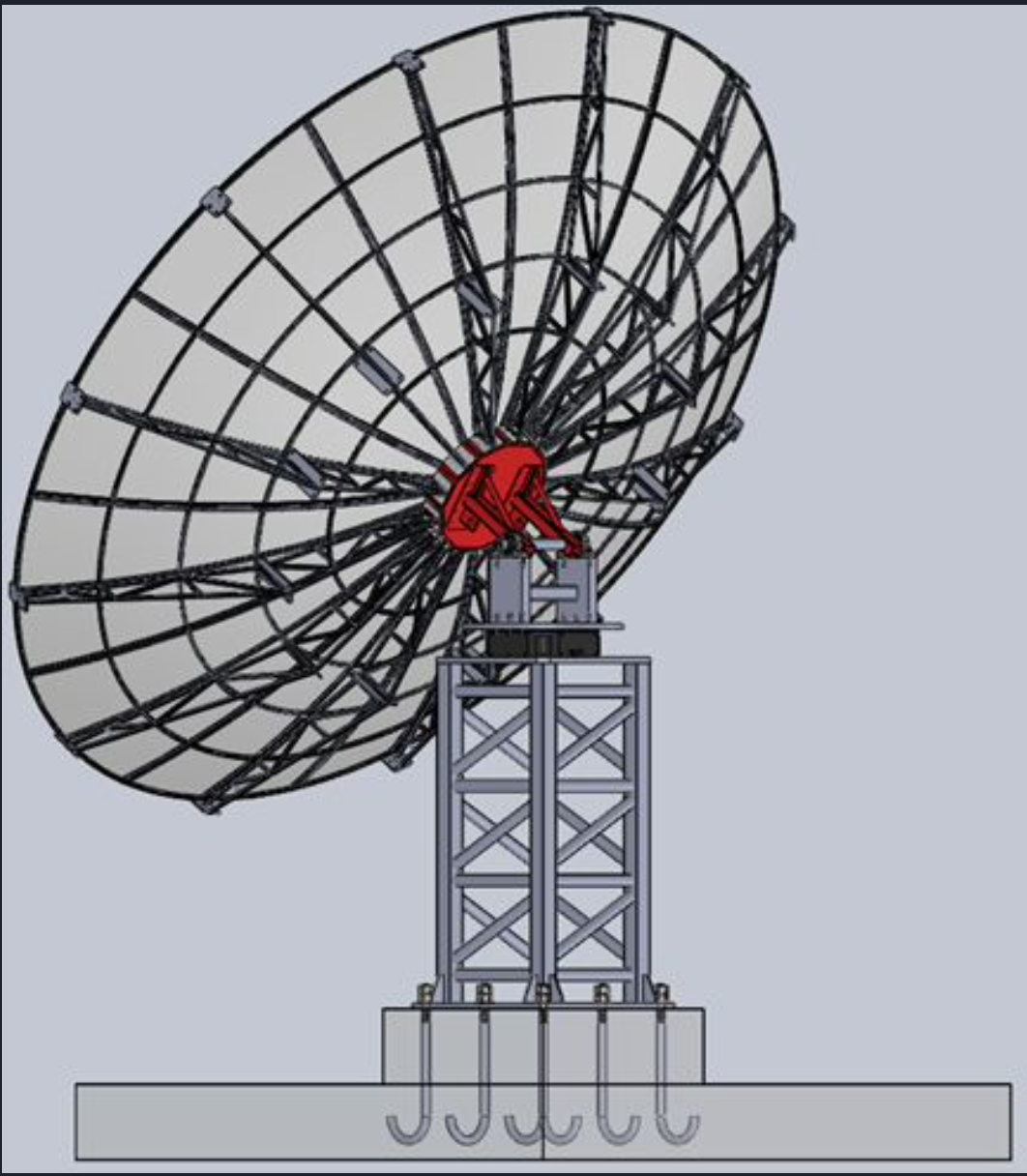}
  \caption{Design of MIA individual telescopes.}
  \label{fig:M1}
\end{figure}

\begin{figure}[!t]
  \includegraphics[width=\columnwidth]{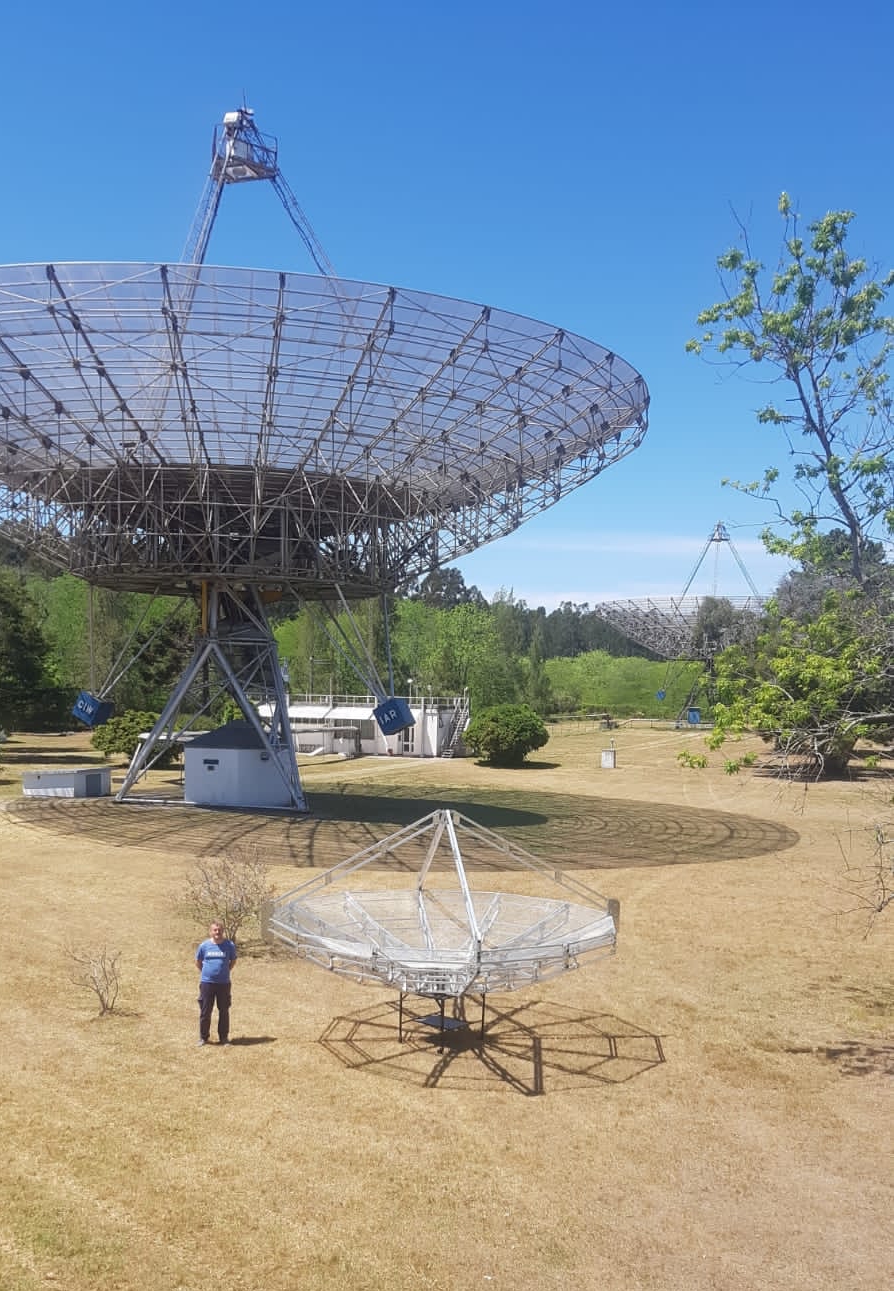}
  \caption{First dish of MIA pathfinder, 2022.}
  \label{fig:M2}
\end{figure}

In addition to these and other upgrades, it is clear that the IAR needs a new observing facility if it is to remain competitive. Because of the severe radio interference at cm wavelength in the province of Buenos Aires, such a facility should be installed in an isolated area and operated remotely.  The project designed to fulfill this need was initiated in 2019 and is called the Multipurpose Interferometric Array (MIA). MIA will be a low frequency (between 50 MHz and 2 GHz) and high bandwidth (250 MHz in phase I) interferometer formed by a core of 16 antennas of 5 m diameter (Fig. \ref{fig:M1}). The maximum baseline of the instrument will be 55 km. It will be located in a site with low radio interference, in the western part of the Argentine Republic. In a second phase (Phase II), the antenna array can be extended up to 64 antennas, including longer baselines. This instrument will allow the achievement of several scientific and technological objectives. MIA will enhance the scientific and technological capabilities of the IAR and will keep the Institute scientifically competitive from an observational point of view until the middle of the 21st century.

The instrument will be able to operate in three main modes: continuum observations, line observations, and high temporal resolution timing measurements. Its main contributions will be in three areas: Galactic variable objects, HI at different redshifts, and fast radio transients.

A prototype called ``MIA Pathfinder'' is now under construction at IAR.  It consists of 1 NODE with 3 antennas (the Phase 1 instrument will have 3 NODES).  It will operate in the L-band (1 - 2.3 GHz) with two orthogonal linear polarizations. The instantaneous bandwidth will be 250 MHz and the receiver temperature will be about 45 K. As of November 2022, the first dish has been completed (Fig. \ref{fig:M2}).

\begin{figure}
  \includegraphics[width=\columnwidth]{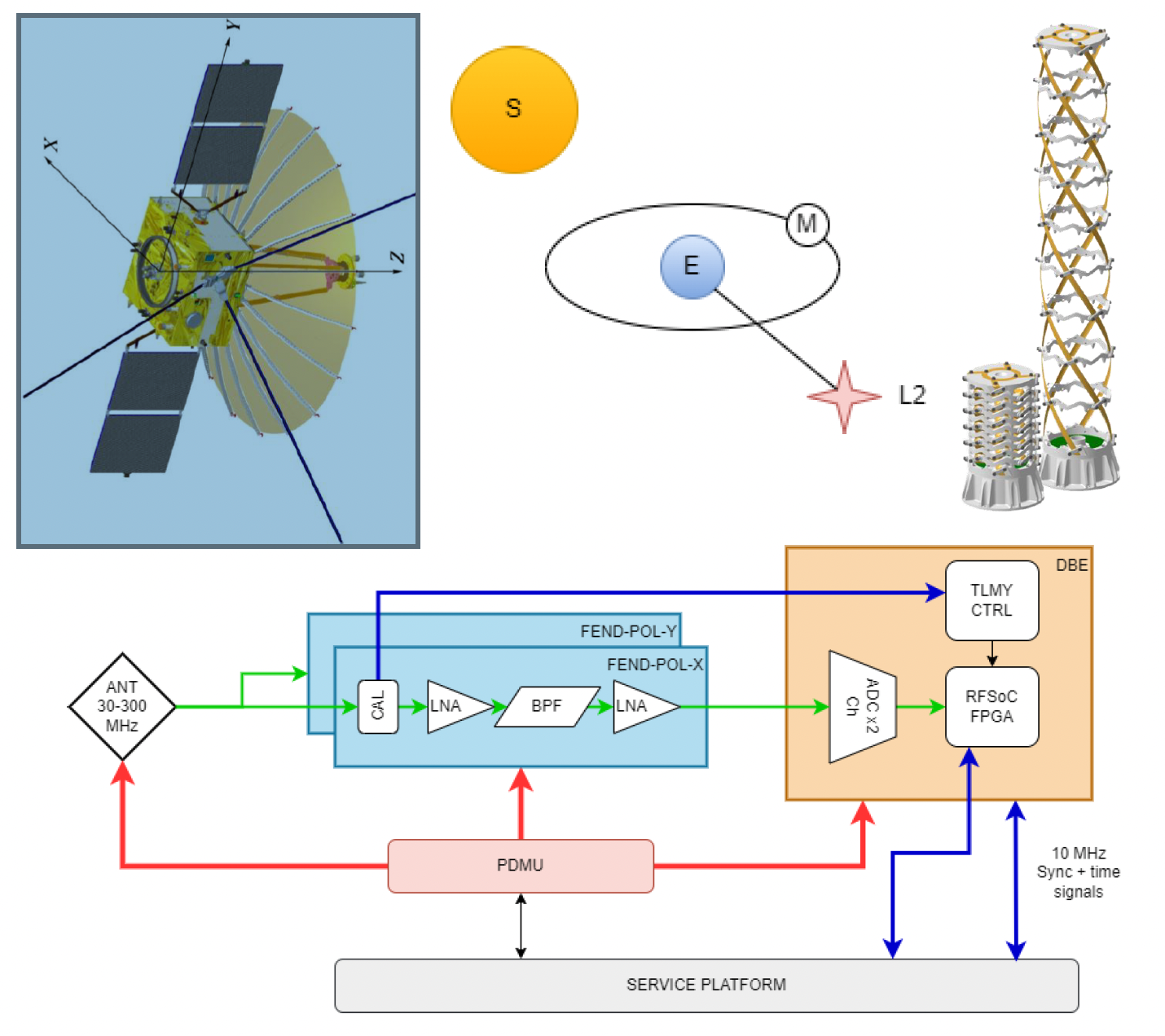}
  \caption{Concept of the Lunar Antenna for Radio Astronomy (LARA).}
  \label{fig:LARA}
\end{figure}

In addition to the new interferometer, the Institute is working on the concept of a lunar radio telescope: the Lunar Antenna for Radio Astronomy (LARA). This instrument is being developed for the Argentine Space Agency CONAE and is a small payload to be transported on an L2 lunar orbiter based on the Netherlands-China Low-Frequency Explorer (NCLE) and CubeRRT. It consists of a deployable antenna using elastic energy and a digital receiver operating in the range of 30 MHz to 300 MHz (Fig. \ref{fig:LARA}). The scientific objective is to provide CONAE with a technological demonstration for a lunar mission that will be able to measure synchrotron radiation from Jupiter, solar bursts, fast radio transients from outside the solar system, and the level of radio interference in the Moon.

\begin{figure}[t!]
  \centering
\includegraphics[width=0.450\textwidth]{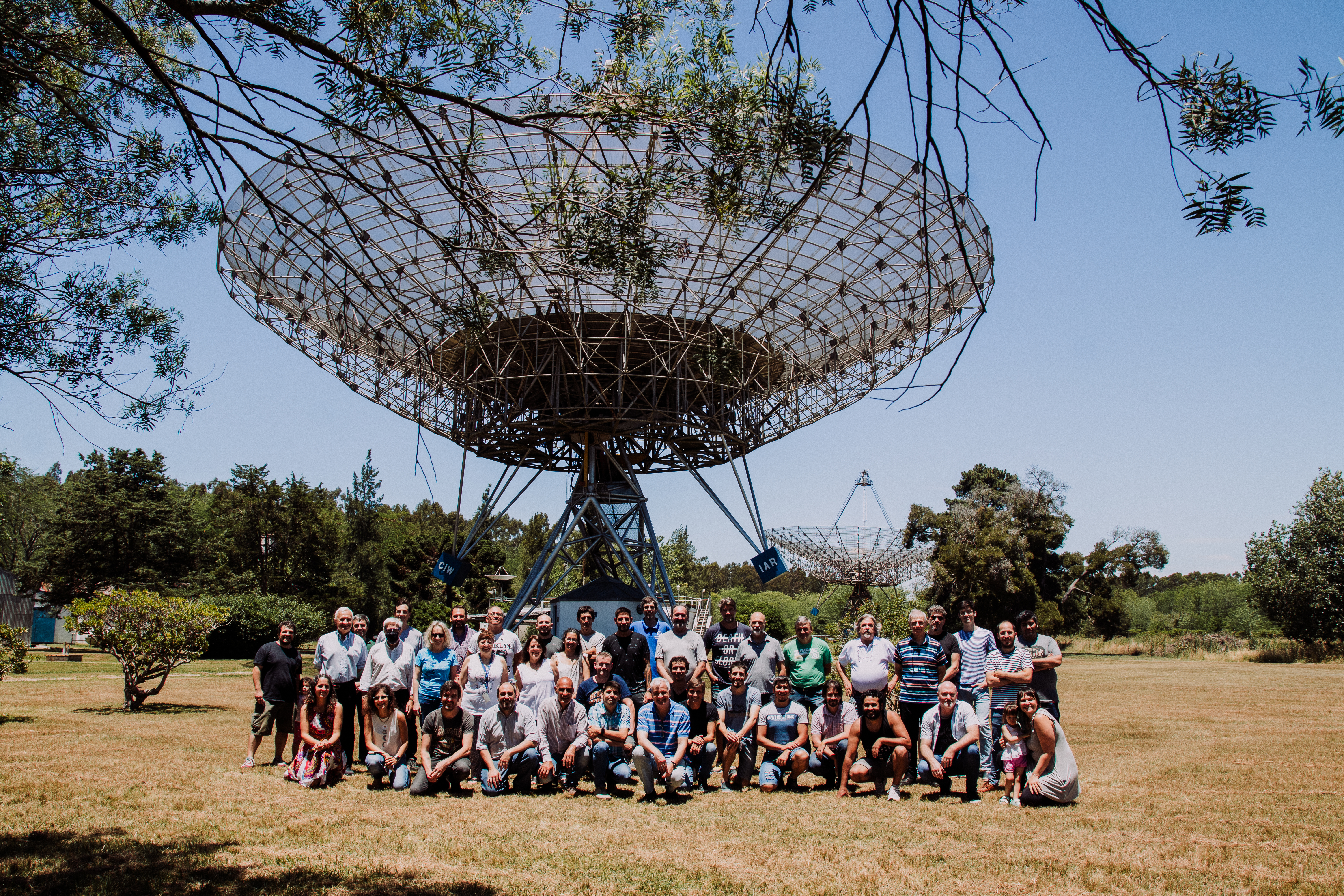}
  \caption{The staff of IAR, 2022.}
  \label{fig:Starff2021}
\end{figure}

\section{Conclusions}

For 60 years, the IAR has been the leading institution in radio astronomy in Argentina. In both good and difficult times, the IAR and its staff have contributed to the development of science and technology in the country and the region. The Institute has evolved from a simple station around a radio telescope in the early 1960s to a center for advanced research and an example of technology development and transfer today: the knowledge gained from the development of astronomical facilities has been applied to the production of a wide variety of technological systems with applications ranging from space research to medical devices. At the same time, the IAR has become a benchmark in the training of human resources, both in astrophysics and engineering. Undoubtedly, it will continue to lead the advancement of knowledge and technology in Argentina for decades to come.

\vspace{0.3cm}

{\bf Acknowledgements:}  I am grateful to Daniela Pérez for a reading the manuscript and making valuable suggestions. I have been supported by grant PIP 11220200100554CO (CONICET) and the State Agency for Research of the Spanish Ministry of Science and Innovation under grant
PID2019-105510GB-C31/AEI/10.13039/501100011033 and through the ``Unit of Excellence María de Maeztu 2020-2023'' award to the Institute of Cosmos Sciences (CEX2019-000918-M). I thank Paula Benaglia, Ileana Andruchow, and the LOC of the meeting for an excellent job. My deepest gratitude to the staff of IAR for their continuous work and efforts that gives life to such a great institution.

\end{document}